\documentclass{aa}  

\usepackage{graphicx}
\usepackage{txfonts}
\usepackage[breaklinks=true,backref=false]{hyperref}
\usepackage{breakurl}
\hypersetup{
    bookmarksopen=false,
    bookmarksnumbered=false,
    unicode=true,           % non-Latin characters in Acrobat’s bookmarks
    pdftoolbar=true,        % show Acrobat’s toolbar?
    pdfmenubar=true,        % show Acrobat’s menu?
    pdffitwindow=false,     % window fit to page when opened
    pdfstartview={FitH},    % fits the width of the page to the window
    pdfnewwindow=true,      % links in new window
    colorlinks=true,       % false: boxed links; true: colored links
    linkcolor=red,          % color of internal links
    citecolor=magenta,        % color of links to bibliography
    filecolor=green,      % color of file links
    urlcolor=blue           % color of external links
}

\usepackage{amssymb}% http://ctan.org/pkg/amssymb
\usepackage{pifont}% http://ctan.org/pkg/pifont
\newcommand{\cmark}{\ding{51}}%
\usepackage{amsmath}
\usepackage{multirow}

\begin{document}

   \title{The Gaia FGK benchmark stars
       \thanks{Based on NARVAL and HARPS data obtained within the Gaia DPAC (Data Processing and Analysis Consortium) and coordinated by the GBOG (Ground-Based Observations for Gaia) working group, and on data retrieved from the ESO-ADP database.} \fnmsep 
       \thanks{The library is available in electronic form at the CDS via anonymous ftp to cdsarc.u-strasbg.fr or via \url{http://cdsweb.u-strasbg.fr/cgi-bin/qcat?J/A+A/}, and at \url{http://www.blancocuaresma.com/s/} } 
   }

   \subtitle{High resolution spectral library}

   \author{ S. Blanco-Cuaresma \inst{1, 2}
            \and C. Soubiran \inst{1, 2}
            \and P. Jofr\'e \inst{1, 2, 3}
            \and U. Heiter \inst{4} 
          }
   \offprints{S. Blanco-Cuaresma, \email{blanco@obs.u-bordeaux1.fr}}

   \institute{Univ. Bordeaux, LAB, UMR 5804, F-33270, Floirac, France.
         \and
            CNRS, LAB, UMR 5804, F-33270, Floirac, France
         \and
            Institute of Astronomy, University of Cambridge, Madingley Road, Cambridge CB3 0HA, U.K.
         \and
            Department of Physics and Astronomy,  Uppsala University, Box 516, 75120 Uppsala, Sweden
            }

   %\date{Received August 31, 2013; accepted September, 2013}

% \abstract{}{}{}{}{} 
% 5 {} token are mandatory
 
  \abstract
   {An increasing number of high-resolution stellar spectra is available today thanks to many past and ongoing spectroscopic surveys. Consequently, numerous methods have been developed to perform an automatic spectral analysis on a massive amount of data. When reviewing published results, biases arise and they need to be addressed and minimized.}
   {We are providing a homogeneous library with a common set of calibration stars (known as the Gaia FGK benchmark stars) that will allow us to assess stellar analysis methods and calibrate spectroscopic surveys.}
   {High-resolution and signal-to-noise spectra were compiled from different instruments. We developed an automatic process to homogenize the observed data and assess the quality of the resulting library.}
   {We built a high-quality library that will facilitate the assessment of spectral analyses and the calibration of present and future spectroscopic surveys. The automation of the process minimizes the human subjectivity and ensures reproducibility. Additionally, it allows us to quickly adapt the library to specific needs that can arise from future spectroscopic analyses. }
   {}


   \maketitle
%
%________________________________________________________________

%\todototoc
%\listoftodos[TODO/Pending tasks]

\section{Introduction}

Investigations into how the Milky Way is formed and its evolution are being revolutionized thanks to the many ongoing stellar spectroscopic surveys such as  SDSS \citep{2000AJ....120.1579Y}, LAMOST \citep{2006ChJAA...6..265Z}, RAVE \citep{2006AJ....132.1645S}, Gaia \citep{2001A&A...369..339P}, Gaia-ESO \citep[GES,][]{2012Msngr.147...25G}, HERMES/GALAH \citep{2010gama.conf..319F}  and APOGEE \citep{2008AN....329.1018A}. Tracing the chemical and dynamical signatures of large samples of stars  helps us to distinguish the different Galactic components and thus understand when and how the different Galactic formation scenarios took place. The quantity of spectroscopic data available today requires the development of automatic spectral analysis. Numerous methods have been developed over the past years \citep[e.g.,,][to name a few]{1996A&AS..118..595V, 1998A&A...338..151K, 2006MNRAS.370..141R, 2008AJ....136.2022L, 2009A&A...501.1269K, 2010A&A...517A..57J, 2012A&A...544A.154P, 2013ApJ...766...78M, 2013A&A...558A..38M} to asses large datasets, where each of them have different approaches to calibrate and evaluate their results. 

However, each survey has its own setup (e.g., spectral range, resolution) and each spectral analysis code has its own particularities (i.e., continuum normalization, atomic line lists). The consequence is that the resulting parameters cannot be directly combined and used for galactic and stellar studies. Thus, spectroscopic calibration with a common reference set of stars is required.

There are several stellar spectral libraries available in the community that are used for calibration in some sense \cite[see, e.g.,][for a compilation]{2005MSAIS...8..170M}, providing a large sample of good spectra of stars covering a large part of the Hertzsprung-Russel (HR) diagram and metallicities. Examples of them are ELODIE \citep{2001A&A...369.1048P}, Indo-US \citep{2004ApJS..152..251V}, MILES \citep{2006MNRAS.371..703S}, StarCAT \citep{2010ApJS..187..149A}, and UVES-POP \citep{2003Msngr.114...10B}. These libraries contain a large number of stars (usually above 1,000) and they differ from each other in terms of resolution and wavelength coverage. They are frequently used for stellar population synthesis models and galactic studies \citep[e.g.,][and references therein]{2012MNRAS.424..157V, 2009ApJ...690..427P, 2005MNRAS.364..503Z} and for calibration or validation of methods that determine stellar parameters from stellar spectra \citep[e.g][]{2008AJ....136.2070A, 2009A&A...501.1269K, 2011RAA....11..924W}. Nevertheless, the Sun is frequently the only calibration star in common between different methods/surveys and, depending on the survey, its observation is not always possible.

%\newpage % To avoid: pdfTeX error (ext4): \pdfendlink ended up in different nesting level than \pdfstartlink

Our motivation for defining the Gaia FGK benchmark stars is to provide a common set of calibration stars beyond the Sun, covering different regions of the HR diagram and spanning a wide range in metallicity. They will be used as pillars for the calibration of the parameters that will be derived for one billion stars by Gaia \citep{2001A&A...369..339P}. The defining property of these stars is that we know their radius and bolometric flux, which allows us to estimate their effective temperature and surface gravity {\it fundamentally}, namely, independent of the spectra. In Heiter et al (in prep, Paper~I), we provide the main properties of our sample of the Gaia FGK benchmark stars and describe the determination of temperature and gravity. In this article (Paper~II), we introduce the spectral library of the Gaia FGK benchmark stars. In \citet[][Paper~III]{2013arXiv1309.1099J}, we analyse our library with the aim to provide a homogeneous scale for the metallicity. 

The current sample of Gaia benchmark stars is composed of bright, well-known FGK dwarfs, subgiants, and giants with metallicities between solar and $-2.7$~dex. We selected stars for which angular diameter and bolometric flux measurements are available or possible. They have accurate parallax measurements, mostly from the HIPPARCOS mission. The sample contains several visual binary stars. In particular, both the A and B components are included for the $\alpha$ Cen and the 61 Cyg systems. The star $\eta$ Boo is a single-lined spectroscopic binary \citep{2005A&A...436..253T}.

The fastest rotators in the sample are $\eta$ Boo and HD 49933, with $v\sin i \gtrsim 10$~km~s$^{-1}$. The metal-poor dwarf Gmb 1830 has the highest proper motion (4.0 and $-$5.8 arcsec/yr in right ascension and declination, respectively). Most of the other stars have proper motions less than 1 arcsec/yr.

Our library provides a homogeneous set of high-resolution and high signal-to-noise ratio (S/N) spectra for the 34 benchmark stars. Moreover, the stellar parameters of these benchmark stars were determined consistently and homogeneously, making them perfect for being used as reference. This library of 34 benchmark stars is therefore a powerful  tool to cross-calibrate methods and stellar surveys, which is crucial for having a better understanding of the structure and evolution of the Milky Way. 

The observed spectra of the benchmark stars were obtained from different telescopes with different instruments and specifications (i.e., resolution and sampling). We developed an automatic process to transform the spectra into one final homogeneous dataset. This allows us to easily generate new versions of the library adapted to the needs of specific surveys (i.e., downgrading the resolution or selecting a different spectral region). Additionally, since reproducibility is one of the main pillars of science, our code will be provided under an open source license to any third party wishes to reproduce the results \citep{2013arXiv1312.4545B}.

This article is structured as follows. In Sect.~\ref{sub:observational_data}, we describe the original observed spectra and its sources, while in Sect.~\ref{sub:data_handling_and_processing} we introduce the computer process that was developed to create the library. Section~\ref{sub:validation} presents the different tests that were performed to validate the correctness of processing and the consistency of the library. In Sect.~\ref{sub:results}, we describe the resulting library's elements that we provide, and finally, we conclude the paper in Sect.~\ref{sub:conclusions}

\section{Observational data}\label{sub:observational_data}

The original observed spectra come mainly from the archives of three different instruments (NARVAL, HARPS, and UVES). In some cases, observations of the same star were obtained by different telescopes, which gives us the possibility to evaluate instrumental effects (see Table ~\ref{tab:summary_data} for a general overview).

\begin{table}
\caption{List of the high-resolution spectra available per benchmark star and instrument.}\label{tab:summary_data}
\begin{center}
\tabcolsep=0.11cm
\begin{tabular}{ |c |c| c| c| c| c|}
\hline 
 Star      &   NARVAL &  HARPS & UVES &  UVES-POP\\
\hline
\hline
$\alpha$ Cen A &   & \cmark & \cmark &  \\
\hline
$\alpha$ Cen B &   & \cmark &   &  \\
\hline
$\alpha$ Cet & \cmark & \cmark & \cmark &  \\
\hline
$\alpha$ Tau & \cmark & \cmark &   &  \\
\hline
$\beta$ Ara &   & \cmark &   &  \\
\hline
$\beta$ Gem &   & \cmark &   &  \\
\hline
$\beta$ Hyi &   & \cmark & \cmark & \cmark\\
\hline
$\beta$ Vir & \cmark & \cmark &   &  \\
\hline
$\delta$ Eri & \cmark & \cmark & \cmark & \cmark\\
\hline
$\epsilon$ Eri &   & \cmark & \cmark & \cmark\\
\hline
$\epsilon$ For &   & \cmark &   &  \\
\hline
$\epsilon$ Vir & \cmark & \cmark &   &  \\
\hline
$\eta$ Boo & \cmark & \cmark &   &  \\
\hline
$\gamma$ Sge & \cmark &   &   &  \\
\hline
$\mu$ Ara &   & \cmark & \cmark &  \\
\hline
$\mu$ Cas & \cmark &   &   &  \\
\hline
$\mu$ Leo & \cmark &   &   &  \\
\hline
$\psi$ Phe &   & \cmark &   &  \\
\hline
$\tau$ Cet & \cmark & \cmark &   &  \\
\hline
$\xi$ Hya &   & \cmark &   &  \\
\hline
18 Sco & \cmark & \cmark &   &  \\
\hline
61 Cyg A & \cmark &   &   &  \\
\hline
61 Cyg B & \cmark &   &   &  \\
\hline
Arcturus & \cmark & \cmark & \cmark & \cmark\\
\hline
Gmb 1830 & \cmark &   &   &  \\
\hline
HD 107328 & \cmark & \cmark &   &  \\
\hline
HD 122563 & \cmark & \cmark & \cmark & \cmark\\
\hline
HD 140283 & \cmark & \cmark & \cmark & \cmark\\
\hline
HD 220009 & \cmark & \cmark &   &  \\
\hline
HD 22879 & \cmark & \cmark &   &  \\
\hline
HD 49933 &   & \cmark &   &  \\
\hline
HD 84937 & \cmark & \cmark & \cmark & \cmark\\
\hline
Procyon & \cmark & \cmark & \cmark & \cmark\\
\hline
Sun & \cmark & \cmark & \cmark &  \\
\hline
\end{tabular}
\end{center}
\label{tab:bs_library}
\end{table}%

\subsection{NARVAL spectra}
The NARVAL spectropolarimeter is mounted on the 2m Telescope Bernard Lyot \citep{2003EAS.....9..105A} located at Pic du Midi (France). The data from NARVAL were reduced with the Libre-ESpRIT pipeline \citep{donati97}. Most of these spectra were taken within a large programme proposed as part of the ``Ground-based observations for Gaia" (P.I: C. Soubiran). The benchmark stars observed with this instrument are listed in Table ~\ref{tab:narval}, where we indicate the S/N and the radial velocity.

Note that one of the solar spectra was created by co-adding 11 spectra of asteroids with the aim to have higher S/N. The asteroids were observed on different nights, therefore there is no observation date in Table ~\ref{tab:narval}. Another solar spectrum corresponding to one single asteroid observation (Metis) with low S/N is included in our sample, which can be used to study S/N effects in spectral analysis.

NARVAL spectra cover a large wavelength range ($\sim 300 - 1100$~nm), with a resolving power \footnote{The terms "resolving power" and "resolution" refer to the relation $R = \frac{\lambda}{\Delta\lambda}$, although we prefer to use the former when talking about instrumental capabilities and the latter for already observed spectra} that varies for different observation dates and along the wavelength range, typically from $75,000$ around 400~nm to $85,000$ around 800~nm. However, it is acceptable to initially assume a constant resolving power of R$\simeq$81,000 as we prove in Sect.~\ref{sub:resolution}.

\begin{table}
\begin{center}
\caption{Spectra observed with the NARVAL spectrograph (average resolving power of $\sim$81,000). } 
\label{tab:narval}
\tabcolsep=0.11cm
\begin{tabular}{c c c c}
\hline
\textbf{Star} & \textbf{ S/N } & \textbf{RV} & \textbf{Date} \\
\hline
18 Sco & 310 / 393 / 429 & 11.62 $\pm$ 0.05 & 2012-03-10\\
61 Cyg A & 248 / 375 / 429 & -65.84 $\pm$ 0.04 & 2009-10-16\\
61 Cyg B & 290 / 464 / 548 & -64.71 $\pm$ 0.05 & 2009-10-13\\
$\alpha$ Cet & 192 / 296 / 367 & -26.12 $\pm$ 0.04 & 2009-12-09\\
$\alpha$ Tau & 209 / 319 / 382 & 54.31 $\pm$ 0.04 & 2009-10-26\\
Arcturus & 283 / 388 / 443 & -5.31 $\pm$ 0.04 & 2009-12-11\\
$\beta$ Vir & 349 / 414 / 437 & 4.39 $\pm$ 0.06 & 2012-01-09\\
$\delta$ Eri & 277 / 356 / 393 & -6.27 $\pm$ 0.03 & 2009-10-26\\
$\epsilon$ Vir & 309 / 388 / 425 & -14.37 $\pm$ 0.04 & 2009-11-27\\
$\eta$ Boo & 366 / 433 / 452 & -6.04 $\pm$ 0.12 & 2009-12-11\\
$\gamma$ Sge & 301 / 467 / 565 & -34.53 $\pm$ 0.03 & 2011-09-30\\
Gmb 1830 & 334 / 420 / 458 & -98.22 $\pm$ 0.07 & 2012-01-09\\
HD 107328 & 278 / 384 / 439 & 36.41 $\pm$ 0.04 & 2009-11-26\\
HD 122563 & 274 / 352 / 398 & -26.09 $\pm$ 0.18 & 2009-11-27\\
HD 140283 & 265 / 317 / 345 & -170.56 $\pm$ 0.44 & 2012-01-09\\
HD 220009 & 278 / 384 / 441 & 40.36 $\pm$ 0.04 & 2009-10-16\\
HD 22879 & 256 / 306 / 326 & 120.37 $\pm$ 0.09 & 2009-11-27\\
HD 84937 & 189 / 220 / 231 & -14.89 $\pm$ 0.51 & 2012-01-08\\
$\mu$ Cas & 220 / 278 / 302 & -96.48 $\pm$ 0.06 & 2009-11-26\\
$\mu$ Leo & 307 / 415 / 465 & 13.53 $\pm$ 0.03 & 2011-12-10\\
Procyon & 676 / 790 / 824 & -5.75 $\pm$ 0.08 & 2012-03-16\\
Sun (Metis) & 36 / 47 / 52 & 2.87 $\pm$ 0.05 & 2010-04-25\\
Sun (co-added) & 584 / 723 / 778 & 5.41 $\pm$ 0.05 & -\\
$\tau$ Cet & 296 / 368 / 399 & -16.65 $\pm$ 0.05 & 2009-12-08\\
\hline
\end{tabular}
\tablefoot{The S/N ratio is reported for three different segments: 480 - 540 / 540 - 610 / 610 - 680~nm. The measured radial velocities are in km/s. The last column corresponds to the observation date. }
\end{center}
\end{table}

\subsection{HARPS spectra}
\label{sub:HARPS}
HARPS is the ESO facility for the measurement of radial velocities with very high accuracy. It is fibre-fed by the Cassegrain focus of the 3.6m telescope in La Silla \citep{2003Msngr.114...20M}. The spectra were reduced by the HARPS Data Reduction Software (version 3.1). Most of the data for benchmark stars were obtained within the programme for critical tests on stellar atmosphere models within the Gaia-SAM collaboration (P.I: U. Heiter). The remaining HARPS data were taken from the public archives.

\begin{table}
\begin{center}
\caption{Spectra observed with the HARPS instrument (average resolving power of $\sim$115,000). } 
\label{tab:harps}
\tabcolsep=0.08cm
\footnotesize
\begin{tabular}{c c c c c}
\hline
\textbf{Star} & \textbf{ S/N } & \textbf{RV} & \textbf{$\Delta$RV}& \textbf{Date} \\
\hline

18 Sco & 146 / 168 / 173 & 11.82 $\pm$ 0.04 & -0.02 & 2009-05-20\\
$\alpha$ Cen A & 381 / 442 / 487 & -22.6 $\pm$ 0.04 & -0.02 & 2005-04-08\\
$\alpha$ Cen A & 433 / 495 / 540 & -22.6 $\pm$ 0.04 & -0.02 & 2005-04-19\\
$\alpha$ Cen B & 393 / 467 / 507 & -21.86 $\pm$ 0.03 & -0.05 & 2005-04-08\\
Arcturus & 368 / 470 / 554 & -5.19 $\pm$ 0.03 & -0.01 & 2004-07-08\\
$\beta$ Hyi & 381 / 424 / 465 & 23.16 $\pm$ 0.05 & 0.01 & 2005-11-13\\
$\beta$ Vir & 272 / 313 / 329 & 4.55 $\pm$ 0.05 & 0.03 & 2009-04-10\\
$\delta$ Eri & 436 / 521 / 576 & -6.21 $\pm$ 0.03 & -0.04 & 2005-10-23\\
$\epsilon$ Eri & 383 / 474 / 537 & 16.4 $\pm$ 0.04 & -0.05 & 2005-12-28\\
HD 49933 & 293 / 323 / 329 & -12.06 $\pm$ 0.19 & 0.06 & 2011-01-05\\
$\mu$ Ara & 207 / 250 / 275 & -9.35 $\pm$ 0.04 & -0.02 & 2004-06-08\\
Sun-1 (Ceres) & 227 / 253 / 267 & 3.99 $\pm$ 0.04 & 0.07 & 2006-07-16\\
Sun-2 (Gan.) & 312 / 362 / 389 & 6.02 $\pm$ 0.04 & -0.02 & 2007-04-13\\
Sun-3 (Vesta) & 174 / 194 / 201 & 3.77 $\pm$ 0.04 & 0.0 & 2009-12-25\\
Sun (co-added) & 421 / 483 / 514 & 3.97 $\pm$ 0.04 & - & -\\
$\tau$ Cet & 219 / 260 / 282 & -16.57 $\pm$ 0.04 & -0.04 & 2008-09-09\\
$\alpha$ Cet & 165 / 228 / 284 & -25.66 $\pm$ 0.04 & -0.16 & 2007-10-22\\
$\alpha$ Tau & 47 / 69 / 86 & 54.21 $\pm$ 0.03 & -0.06 & 2007-10-22\\
$\beta$ Ara & 285 / 408 / 488 & 0.22 $\pm$ 0.04 & 0.25 & 2007-09-29\\
$\beta$ Gem & 287 / 359 / 416 & 3.42 $\pm$ 0.03 & -0.04 & 2007-11-06\\
$\epsilon$ For & 281 / 330 / 358 & 40.8 $\pm$ 0.04 & -0.01 & 2007-10-22\\
$\epsilon$ Vir & 319 / 386 / 421 & -14.28 $\pm$ 0.03 & -0.02 & 2008-02-24\\
$\eta$ Boo & 358 / 410 / 439 & -2.23 $\pm$ 0.12 & 0.06 & 2008-02-24\\
HD 107328 & 343 / 452 / 524 & 36.66 $\pm$ 0.03 & -0.02 & 2008-02-24\\
HD 122563 & 353 / 430 / 492 & -26.24 $\pm$ 0.17 & 0.07 & 2008-02-24\\
HD 140283 & 441 / 497 / 535 & -170.46 $\pm$ 0.42 & 0.05 & 2008-02-24\\
HD 220009 & 262 / 342 / 398 & 40.18 $\pm$ 0.04 & 0.01 & 2007-10-22\\
HD 22879 & 296 / 322 / 336 & 120.38 $\pm$ 0.08 & -0.03 & 2007-10-22\\
HD 84937 & 444 / 484 / 513 & -14.76 $\pm$ 0.49 & 0.24 & 2007-12-03\\
$\xi$ Hya & 318 / 386 / 422 & -4.53 $\pm$ 0.04 & -0.01 & 2008-02-24\\
Procyon & 352 / 373 / 377 & -3.11 $\pm$ 0.08 & 0.0 & 2007-11-06\\
$\psi$ Phe & 274 / 352 / 460 & 3.07 $\pm$ 0.06 & -0.31 & 2007-09-30\\
\hline
\end{tabular}
\tablefoot{The S/N ratio is reported for three different segments: 480 - 540 / 540 - 610 / 610 - 680~nm. The measured radial velocities are in km/s and $\Delta$RV is the difference from the reported velocity by HARPS pipeline. The last column corresponds to the observation date. }
\end{center}
\end{table}

The list of HARPS spectra can be found in Table ~\ref{tab:harps}. The solar spectra correspond to two observations of asteroids (Ceres and Vesta) and one observation of one of Jupiter's moons (Ganymede). We could obtain them directly from the public archive thanks to \cite{2012A&A...544A.125M}, who presented a detailed analysis of absorption lines of the those spectra. Additionally, we co-added those spectra to have a solar spectrum with higher S/N.

The spectra for the stars HD84937 and HD140283 are the result of the co-addition of four and two individual observed spectra, respectively. The reason for combining these spectra is to increase the S/N. The spectra of each star were taken on the same night, with the date indicated in Table ~\ref{tab:harps}.

The spectral range covered is $378 - 691$~nm, but as the detector consists of a mosaic of two CCDs, one spectral order (from 530~nm to 533~nm) is lost in the gap between the two chips.

\subsection{UVES spectra}
\label{sub:UVES}

The UVES spectrograph is hosted by unit telescope 2 of ESO's VLT \citep{2000SPIE.4008..534D}. We took the spectra of benchmark stars available the Advanced Data Products collection of the ESO Science Archive Facility\footnote{\url{http://archive.eso.org/eso/eso\_archive\_adp.html}}  (reduced by the standard UVES pipeline version 3.2, \citealp{2000Msngr.101...31B}) by selecting only the most convenient\footnote{Those spectra with high S/N and smooth continuum, given that sometimes the merging of the orders did not produce a smooth spectrum along the wavelength range} spectrum for each star based on visual inspection. 

The setup used for each observation (CD\#3, centered around 580~nm) provides a spectrum with two different parts which approximately cover the ranges from 476 to 580~nm (lower part) and from 582 to 683~nm (upper part). The resolution of each spectrum and lower/upper parts might be different depending on the slit width used during the observation. There is a relation between the slit width (or the seeing if it is considerable smaller than the slit width) and the resolution, which depends on the date of observation\footnote{\url{http://www.eso.org/observing/dfo/quality/UVES/reports/HEALTH/\\trend\_report\_ECH\_RESOLUTION\_DHC\_HC.html}}. We went through those relations for each spectrum and estimated its resolution which can be found in Table ~\ref{tab:uves}.

\begin{table}
\begin{center}
\centering          
\caption{Spectra observed with the UVES instrument. } 
\label{tab:uves}
\tabcolsep=0.02cm
\footnotesize
\begin{tabular}{c c c c c}
\hline
\textbf{Star} & \textbf{ S/N } & \textbf{RV} & \textbf{R} & \textbf{Date} \\
\hline
$\alpha$ Cen A & 287 / 328 / 324 & -23.13 $\pm$ 0.04 & 90000 | 85000 & 2000-04-11\\
$\alpha$ Cet & 157 / 209 / 227 & -25.53 $\pm$ 0.04 & 115000 | 90000 & 2003-08-11\\
Arcturus & 213 / 291 / 306 & -5.18 $\pm$ 0.03 & 115000 | 95000 & 2004-08-03\\
$\beta$ Hyi & 407 / 447 / 411 & 22.96 $\pm$ 0.05 & 85000 | 78000 & 2001-07-27\\
$\delta$ Eri & 176 / 208 / 217 & -5.87 $\pm$ 0.03 & 85000 | 75000 & 2001-11-29\\
$\epsilon$ Eri & 204 / 231 / 222 & 15.99 $\pm$ 0.04 & 115000 | 90000 & 2001-10-02\\
HD 122563 & 288 / 326 / 327 & -26.66 $\pm$ 0.17 & 82000 | 72000 & 2002-02-19\\
HD 140283 & 279 / 305 / 298 & -170.79 $\pm$ 0.41 & 85000 | 78000 & 2001-07-09\\
HD 84937 & 215 / 229 / 228 & -15.19 $\pm$ 0.48 & 80000 | 72000 & 2002-11-28\\
$\mu$ Ara & 292 / 327 / 309 & -9.25 $\pm$ 0.04 & 115000 | 95000 & 2003-09-05\\
Procyon & 349 / 379 / 356 & -2.35 $\pm$ 0.08 & 82000 | 75000 & 2002-10-08\\
Sun (Vesta) & 337 / 383 / 418 & -5.52 $\pm$ 0.04 & 78000 | 78000 & 2000-10-18\\
\hline
\end{tabular}
\tablefoot{The S/N ratio is reported for three different segments: 480 - 540 / 540 - 610 / 610 - 680~nm. The measured radial velocities are in km/s. The reported resolving power correspond to the lower and upper parts of each spectrum. The last column corresponds to the observation date. }
\end{center}
\end{table}

\subsection{UVES-POP spectra}

The UVES Paranal Observatory Project UVES-POP library \citep[processed with data reduction tools specifically developed for that project]{2003Msngr.114...10B}
contains stellar spectra along the complete UVES wavelength range (300-1000~nm with two gaps around 580~nm and 860~nm). The benchmark stars that are included in this library are listed in Table ~\ref{tab:uves-pop}.

\begin{table}
\begin{center}
\caption{Spectra observed with the UVES instrument and processed by the UVES-POP pipeline (average resolving power of 80,000).}
\label{tab:uves-pop}
\tabcolsep=0.11cm
\begin{tabular}{c c c c}
\hline
\textbf{Star} & \textbf{ S/N } & \textbf{RV} & \textbf{Date} \\
\hline
Arcturus & 1123 / 1312 / 1235 & -5.48 $\pm$ 0.03 & 2003-02-16\\
$\beta$ Hyi & 596 / 690 / 717 & 22.99 $\pm$ 0.06 & 2001-07-25\\
$\delta$ Eri & 421 / 547 / 606 & -5.94 $\pm$ 0.03 & 2001-11-28\\
$\epsilon$ Eri & 1468 / 1625 / 1808 & 16.09 $\pm$ 0.04 & 2002-10-11\\
HD 122563 & 846 / 934 / 624 & -26.73 $\pm$ 0.18 & 2002-08-18\\
HD 140283 & 807 / 901 / 864 & -170.75 $\pm$ 0.43 & 2001-07-08\\
HD 84937 & 511 / 537 / 567 & -15.13 $\pm$ 0.48 & 2002-11-28\\
Procyon & 1297 / 1449 / 948 & -2.25 $\pm$ 0.08 & 2002-10-07\\
\hline
 \end{tabular}
\tablefoot{The S/N ratio is reported for three different segments: 480 - 540 / 540 - 610 / 610 - 680~nm. The measured radial velocities are in km/s. The last column corresponds to the observation date. }
\end{center}
\end{table}

\subsection{Atlas spectra}

We included the already normalized atlas spectra from \cite{2000vnia.book.....H} for the Sun and Arcturus. The observations were made with the Coude Feed Telescope on Kitt Peak with the spectrograph in the Echelle mode, reaching a resolving power of 150,000 and a high S/N. The authors of that work also removed the telluric lines from the provided spectra.

The inclusion of this observations in the library is only for normalization validation purposes (see Sect. \ref{sub:normalization_validation}); they were not treated by the same homogeneous normalization process as the other spectra.

\section{Data handling and processing}\label{sub:data_handling_and_processing}

An automatic computational process was developed to transform the observed spectra presented in Sect.~\ref{sub:observational_data} into a homogeneous library of the benchmark stars. Since the wavelength range varies from one set of observations to another, we chose to limit the current library to the range between 480 and 680~nm, where all the spectra provide their best S/N. Additionally, the range matches the interests of the Gaia-ESO Survey \citep[GES, ][]{2012Msngr.147...25G} given their UVES setup on the ESO-VLT telescope.

\begin{figure}
    \includegraphics[width=\linewidth, trim = 50mm 30mm 50mm 30mm]{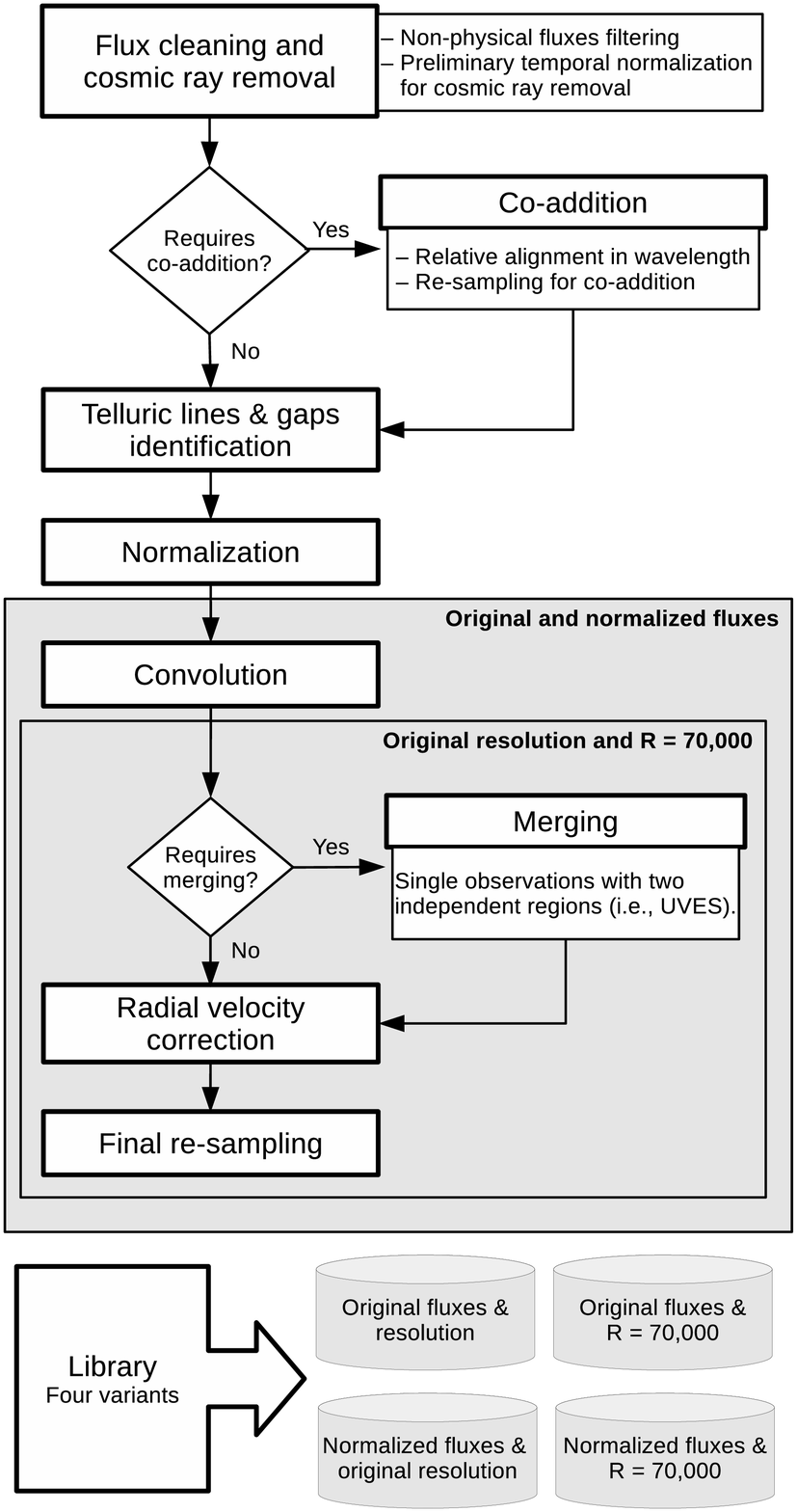}
    \caption{Flowchart of the automatic computational process developed to build the library.}
    \label{fig:process}
\end{figure}

In this section, we describe how the data are treated to determine and correct the radial velocity of the star, fit the continuum and normalize, convolve the spectra to a common desired resolution, and re-sample the spectra to finally obtain a homogeneous library (see Fig.~\ref{fig:process} for a general overview).

\subsection{Cleaning and cosmic rays removal}

The spectra obtained by different instruments have different conventions to indicate errors in the observation. For instance, bad pixels can be marked with negative fluxes. These values should not be considered as good data, therefore we automatically ignore them.

On the other hand, spectra may contain residuals from cosmic rays which can affect the normalization process and future analyses. We implemented an algorithm that detects them automatically using a median filter to smooth out single-measurement deviations. The process is as follows:

\begin{enumerate}
    \item Re-sample the spectrum to have a homogeneous sampling of 0.001~nm steps (small enough to respect the original resolution).
    \item Obtain a smoothed spectrum by applying a median filter with a window of 15 points, which equals to 0.015~nm.
    \item Re-sample the smoothed spectrum to the original wavelength grid again.
    \item Subtract the smoothed spectrum from the original one.
    \item Mark all the points with differences higher than 0.01 (1\% of total flux) which are located above the continuum as cosmic rays (determined by a preliminary execution of the steps described in Sect.~\ref{sub:normalization}).
\end{enumerate}

Cosmic rays may also appear under the continuum, but if we attempt to remove them, we might also affect the upper parts of sharply blended lines. Therefore, we preferred to limit ourselves to cosmic rays above the continuum.

\subsection{Co-addition}
\label{sub:coaddition}

As specified in Sect.~\ref{sub:HARPS}, several spectra observed by HARPS were co-added for the stars HD140283 and HD84937. Additionally, three solar spectra observed by HARPS and 11 asteroids observed by NARVAL were co-added into two single spectra (one per instrument).
                
For each co-addition, we selected one spectrum as reference, we removed those regions that are potentially affected by telluric lines (see Sect.~\ref{sub:tellurics}) and we executed a cross-correlation process (see Sect.~\ref{sub:cross-correlation}) against the other spectra of the same star. Finally, we aligned all the spectra (zero differential velocity), re-sampled them (see Sect.~\ref{sub:resampling}), and summed up their fluxes.

\subsection{Telluric lines}
\label{sub:tellurics}

Telluric lines from Earth's atmosphere contaminate ground-observed spectra and might affect stellar analyses. We identified automatically those potentially affected regions which allows the users to easily ignore or remove them.

In this process, a synthetic spectrum from TAPAS \citep{2013arXiv1311.4169B} was used. TAPAS is an on-line service that provides simulated atmospheric transmission spectra for specific observing conditions.

We do not need to recover the exact telluric spectrum for the day and place of observation since our goal is not to correct them but to identify the potentially affected regions. Therefore, we obtained a synthetic spectrum as if all the observations were done at ESO observatory (La Silla, Chile) pointing to the zenith (angle of zero degrees).

Once the non-convolved synthetic spectrum was obtained and normalized (as described in Sect.~\ref{sub:normalization}), an automatic process for absorption line detection was applied. The process performs the following steps:

\begin{enumerate}
    \item Search local minimum (representing absorption peaks) and maximum points (corresponding to the limits of the absorption line).
    \item Discard lines found that are false positive or noise by ignoring:
        \begin{enumerate}
        \item peaks that have a smaller depth than its nearby maximums (false positives),
        \item peaks too close (2 or less bins away) to the next and previous maximums (noise);
        \end{enumerate}
    \item Discard those line candidates that do not have a minimum depth (1.0\% of depth with respect to the continuum).
    \item Fit a Gaussian model to the line candidates.
    \item Discard lines with bad fits.
\end{enumerate}

The process provides us with the position of the telluric lines peaks and the full-width at half-maximum (FWHM) of the fitted Gaussian. In order to delimit where the lines start and end, we searched for local maximums $4\times$FWHM around each telluric line's peak or we directly used $2\times$FWHM if no local maximum was found. 

The resulting telluric line list was used as a mask for locating the telluric lines in each observed spectra by using cross-correlation (as the one described in Sect.~\ref{sub:cross-correlation}). The identified regions affected by telluric lines were ignored in the normalization process (Sect.~\ref{sub:normalization}).

\subsection{Gaps}
\label{sub:gaps}

Depending on the instrument, some spectra contain gaps (regions without fluxes or all fluxes set to zero). We identified them automatically to be able to easily ignore those regions.

\subsection{Normalization}
\label{sub:normalization}

The spectra were normalized automatically, reducing biases and inhomogeneities due to subjective criteria. For each observed star, the process uses a synthetic spectrum generated with the MARCS model atmospheres \citep{2008A&A...486..951G}, an atomic line list from VALD \citep{2011BaltA..20..503K}, the SPECTRUM code \citep{1994AJ....107..742G} and the atmospheric parameters listed in Table ~\ref{tab:synth_parameters}. The fitting algorithm is as follows:

\begin{enumerate}
    \item Ignore all the fluxes that have a value below $0.98$ in their respective synthetic spectra (computed with the reference atmospheric parameters). This way, we reduce the effect of strong lines in the normalization process.
    \item Ignore gaps and regions affected by telluric lines.
    \item Reduce the noise effects by applying a median filter with a window of $0.01 \mathrm{nm}$ (window sizes were selected after several validation and optimization tests).
    \item Apply a maximum filter with a window of $1.0 \mathrm{nm}$ to select those fluxes that have more probabilities belonging to the continuum.
    \item Fit second degree splines every $1.0 \mathrm{nm}$ to the filtered points and divide the original observed spectrum by the fitted model.
\end{enumerate}

After several tests, we found this was the most robust strategy to homogeneously normalize the library's spectra.

\begin{table}
\caption{Parameters used for the spectral synthesis in the normalization process.}
\label{tab:synth_parameters}
\begin{center}
\tabcolsep=0.11cm
\begin{tabular}{c c c c c c c}
\hline 
\textbf{Star} & \textbf{Teff} & \textbf{log(g)} & \textbf{[M/H]} & \textbf{Vmic} & \textbf{Vmac} & \textbf{v sin(i)} \\
\hline 
$\psi$ Phe & 3472 & 0.51 & -1.23 & 1.54 & 6.26 & 3.00 \\                         
$\alpha$ Cet & 3796 & 0.68 & -0.45 & 1.36 & 5.68 & 3.00 \\                       
$\gamma$ Sge & 3807 & 1.05 & -0.16 & 1.43 & 5.01 & 6.00 \\                       
$\alpha$ Tau & 3927 & 1.11 & -0.37 & 1.36 & 5.23 & 5.00 \\                       
61 Cyg B & 4044 & 4.67 & -0.38 & 0.85 & 5.00 & 1.70 \\                           
$\beta$ Ara & 4173 & 1.04 & -0.05 & 1.24 & 5.09 & 5.40 \\                        
HD 220009 & 4275 & 1.47 & -0.75 & 1.23 & 5.39 & 1.00 \\                          
Arcturus & 4286 & 1.64 & -0.53 & 1.25 & 5.05 & 3.80 \\                           
61 Cyg A & 4374 & 4.63 & -0.33 & 0.86 & 4.19 & 0.00 \\                           
$\mu$ Leo & 4474 & 2.51 & 0.26 & 1.28 & 3.63 & 5.10 \\                           
HD 107328 & 4496 & 2.09 & -0.34 & 1.23 & 4.60 & 1.90 \\                          
HD 122563 & 4587 & 1.61 & -2.74 & 1.13 & 6.13 & 5.00 \\                          
Gmb 1830 & 4827 & 4.60 & -1.46 & 0.82 & 3.49 & 0.50 \\                           
$\beta$ Gem & 4858 & 2.90 & 0.12 & 1.22 & 3.68 & 2.00 \\                         
$\delta$ Eri & 4954 & 3.75 & 0.06 & 1.00 & 3.60 & 0.70 \\                        
$\epsilon$ Vir & 4983 & 2.77 & 0.13 & 1.23 & 3.78 & 2.00 \\                      
$\xi$ Hya & 5044 & 2.87 & 0.14 & 1.23 & 3.72 & 2.40 \\                           
$\epsilon$ Eri & 5076 & 4.60 & -0.10 & 0.88 & 3.22 & 2.40 \\                     
$\epsilon$ For & 5123 & 3.52 & -0.62 & 1.02 & 4.05 & 4.20 \\                     
$\alpha$ Cen B & 5231 & 4.53 & 0.22 & 0.93 & 2.79 & 1.00 \\                      
$\mu$ Cas & 5308 & 4.41 & -0.82 & 0.91 & 3.85 & 0.00 \\                          
$\tau$ Cet & 5414 & 4.49 & -0.50 & 0.94 & 3.70 & 0.40 \\                         
HD 140283 & 5514 & 3.57 & -2.43 & 1.05 & 5.18 & 5.00 \\                          
Sun & 5777 & 4.44 & 0.02 & 1.07 & 4.19 & 1.60 \\                                 
$\alpha$ Cen A & 5792 & 4.30 & 0.24 & 1.11 & 4.01 & 1.90 \\                      
18 Sco & 5810 & 4.44 & 0.01 & 1.08 & 4.34 & 2.20 \\                              
HD 22879 & 5868 & 4.27 & -0.88 & 1.09 & 5.45 & 4.40 \\                           
$\beta$ Hyi & 5873 & 3.98 & -0.07 & 1.16 & 4.82 & 3.30 \\                        
$\mu$ Ara & 5902 & 4.30 & 0.33 & 1.15 & 4.40 & 2.20 \\                           
$\beta$ Vir & 6083 & 4.10 & 0.21 & 1.24 & 5.62 & 2.00 \\                         
$\eta$ Boo & 6099 & 3.80 & 0.30 & 1.30 & 5.74 & 12.70 \\                         
HD 84937 & 6356 & 4.15 & -2.09 & 1.29 & 9.24 & 5.20 \\                           
Procyon & 6554 & 3.99 & -0.04 & 1.48 & 9.71 & 2.80 \\                            
HD 49933 & 6635 & 4.20 & -0.46 & 1.48 & 10.87 & 10.00 \\                         
\hline
\end{tabular}
\tablefoot{The effective temperature (K) and surface gravity (dex) were obtained from Paper~I, the metallicity corresponds to the Fe I LTE abundances (dex) from Paper~III (SPECTRUM code assumes Local Thermodynamic Equilibrium, LTE), the micro/macroturbulence (km/s) is derived from an empirical relation calibrated by the Gaia ESO Survey working groups (M. Bergemman and V. Hill, private communication) and the rotation (km/s) found in the literature (see Papers~I/III).}
\end{center}
\label{tab:bs_library}
\end{table}%

\subsection{Resolution degradation}

The resulting final library was homogenized to the highest minimum resolving power, which corresponds to 70,000. The process convolves the spectra by performing the following steps for each flux value:

\begin{enumerate}
    \item Define a window with size $FWHM_\lambda$ which depends on the original and final desired resolution:

\begin{equation}
    \mathrm{FWHM}_{\lambda} = \sqrt{\left(\frac{\lambda}{R_{\mathrm{final}}}\right)^{2} - \left(\frac{\lambda}{R_{\mathrm{initial}}}\right)^{2}} \mathrm{.}
\end{equation}

    \item Build a Gaussian profile $g\left(\lambda_{x}\right)$ using the sigma corresponding to $FWHM_\lambda$ and the wavelength values of a spectral window around the wavelength $\lambda$ that it is going to be convolved:

\begin{equation}
g\left(\lambda_{x}\right) = \frac{1}{\sqrt{2 \pi \sigma_{\lambda}^{2}}} e^{- \frac{\left(\lambda_{x} - \lambda\right)^{2}}{2 \sigma_{\lambda}^{2}}} \mathrm{,}
\end{equation}

where 
$\sigma_{\lambda} = \frac{\mathrm{FWHM}_{\lambda}}{2 \sqrt{2 \log 2}}$.

    \item Normalize the Gaussian profile and multiply it by the original fluxes in the spectral window. The sum of that operation will be the new convolved value for the wavelength $\lambda$:

\begin{equation}
\mathrm{flux}\left(\lambda\right) = \sum_{\mathrm{window}}^{x} \mathrm{flux}(\lambda_{x})  \left(\frac{g(\lambda_{x})}{\sum_{\mathrm{window}}^{x} g(\lambda_{x})} \right) \mathrm{.}
\end{equation}
\end{enumerate}

\subsection{Merge}

The spectra from the UVES instrument were observed with a setup that provides a separate lower and upper spectral part (see Sect.~\ref{sub:UVES}). In these cases, we performed the cleaning, normalization and convolution separately and, afterwards, the resulting spectra were merged.

\subsection{Radial velocity}
\label{sub:radial_velocity}

\subsubsection{Zero point template}

As a zero point we could use, for instance, the solar HARPS spectrum since this instrument provides high precision radial velocity measurements. We preferred to use the co-added solar NARVAL spectrum because it does not contain any gap, however, in order to transform the NARVAL solar spectrum into a reliable zero point template, we removed the regions affected by telluric lines and we cross-correlated it with a solar HARPS spectrum, which was corrected by using the radial velocity reported by HARPS pipeline. Finally, we corrected the NARVAL spectrum with the relative velocity shift found.

\subsubsection{Cross-correlation}
\label{sub:cross-correlation}

The first stage in the radial velocity determination process is the generation of the velocity profile by the cross-match correlation algorithm \citep{2002A&A...388..632P}, which sums the spectrum's fluxes multiplied by a mask/template function 'p':

\begin{equation}
    C(\mathrm{v}) = \sum_{\lambda} \mathrm{p}(\lambda, \mathrm{v}) \cdot \mathrm{flux}(\lambda) \mathrm{,}
\end{equation} 

where v is the velocity.

\begin{enumerate}
\item Create a wavelength grid uniformly spaced in terms of velocity. This means that an increment in position ($x \rightarrow x+1$) supposes a constant velocity increment (velocity step) but a variable wavelength increment. The following formula is used for determining the wavelength ranges (relativistic Doppler effect):

\begin{equation}
    \lambda_{x+1} = \lambda_{x} + \lambda_{x} \left(1 - \sqrt{\frac{1-\frac{\mathrm{v}}{c}}{1+\frac{\mathrm{v}}{c}}}\right) \mathrm{,}
\end{equation} 

where $c$ is the speed of light in vacuum and $\lambda$ the original wavelengths.

\item Calculate the cross-correlation function \citep{1996A&AS..119..373B, 2007AJ....134.1843A} between the spectrum and the specified template by shifting the template from limits (see below).
\end{enumerate}

Once the velocity profile was constructed from the cross-correlation process, the mean velocity is calculated by fitting a second order polynomial near the peak. Additionally, a Gaussian model is fitted (with fixed mean velocity) to determine other complementary parameters such as the FWHM.

The whole process is repeated two times: Firstly a general estimation is obtained by using a velocity step of $5.0$~km/s ($1.0$~km/s if we are cross-correlating with a telluric line mask) with lower and upper limits of $\pm200$~km/s; Secondly, a more precise value is determined by using a step of $0.25$~km/s with lower and upper limits of $\pm 4 \times$FWHM around the first velocity estimation.

The error in the radial velocity determination is calculated by following \cite{2003MNRAS.342.1291Z}:

\begin{equation}
    \sigma_{\mathrm{v}}^{2} = - \left[ N \frac{C^{\prime\prime}(\mathrm{v})}{C(\mathrm{v})}  \frac{C^{2}(\mathrm{v})}{1 - C^{2}(\mathrm{v})} \right]^{-1} \mathrm{,}
\end{equation} 

where $N$ is the number of bins in the spectrum, $C$ is the cross-correlation function, and $C^{\prime\prime}$ is its second derivative.

Finally, the correction is performed by applying the following formula:

\begin{equation}
    \lambda_{corrected} = \lambda \sqrt{\frac{1-\frac{\mathrm{v}}{c}}{1+\frac{\mathrm{v}}{c}}} \mathrm{.}
\end{equation}

\subsection{Re-sampling}
\label{sub:resampling}

The final step in the homogenization process is to sample all the spectra by establishing a constant increment in wavelength (0.001~nm) from 480 to 680~nm. To do so, we implemented a Bessel's Central-Difference interpolation similar to that used in TGMET  \citep{1998A&A...338..151K}.

A quadratic Bessel’s interpolation formula was employed. It makes use of two points before and two points after the value to be interpolated (except where there are not enough, such as at the beginning and end of the spectrum, where a linear interpolation is performed). The formula is as follows:

\begin{equation}
\begin{split}
\mathrm{f}(\lambda) & = f\left(\lambda_{0}\right) + p \left( f\left(\lambda_{1}\right) - f\left(\lambda_{0}\right) \right) \\
& + \left[ \frac{p \left( p - 1 \right)}{4} \right] \left( f\left(\lambda_{2}\right) - f\left(\lambda_{1}\right) - f\left(\lambda_{0}\right) + f\left(\lambda_{-1}\right) \right) \mathrm{,}
\end{split}
\end{equation}

where $p = \frac{\lambda - \lambda_{0}}{\lambda_{1} - \lambda_{0}}$,  $f(\lambda)$ is the flux, $\lambda$ is the target wavelength, and $\lambda_{-1} < \lambda_{0} < \lambda < \lambda_{1} < \lambda_{2}$. The zero and first order terms correspond to a linear interpolation, while the second order term is a correction factor to that linear interpolation.

\subsection{Errors}
\label{sub:errors}

All the observed spectra have individual errors associated with each measured flux except the atlas spectra. For the latter, we estimated the errors dividing the fluxes by the S/N, which was obtained by calculating the ratio between the mean flux and the standard deviation for groups of ten measurements around each wavelength point and selecting the mean value.

For the operations that implied flux modification (e.g., convolution, continuum normalization), the errors were taken into account and appropriately propagated.

\section{Validation}
\label{sub:validation}

The resulting library was evaluated to guarantee that the data were properly treated and that the spectra present a high level of quality.

\subsection{Normalization}
\label{sub:normalization_validation}

The normalization process should produce similar spectra for the same stars independent of the instrument used for the observation. To validate that statement and therefore the internal coherence, we compared spectra from different instruments by calculating their root mean square (RMS) difference in flux as:

\begin{equation}
    \mathrm{RMS} = \sqrt{\frac{\sum{\left(\mathrm{flux}_{\mathrm{reference}}-\mathrm{flux}\right)^2}}{\mathrm{num\_fluxes}}} \mathrm{,}
\end{equation}

where $\mathrm{flux}_{\mathrm{reference}}$ is the flux from a given spectrum (i.e., the first in the treatment chain).

The test was performed using the whole wavelength range and three individual regions: H-$\alpha$ (653 - 660~nm), H-$\beta$ (483 - 489~nm), and Mg triplet (515 - 520~nm). Regions affected by telluric lines or gaps were not considered in the comparison.

\begin{figure*}
    \begin{centering}
        \includegraphics[width=9cm, scale=0.40,trim = 10mm 10mm 10mm 10mm, clip]{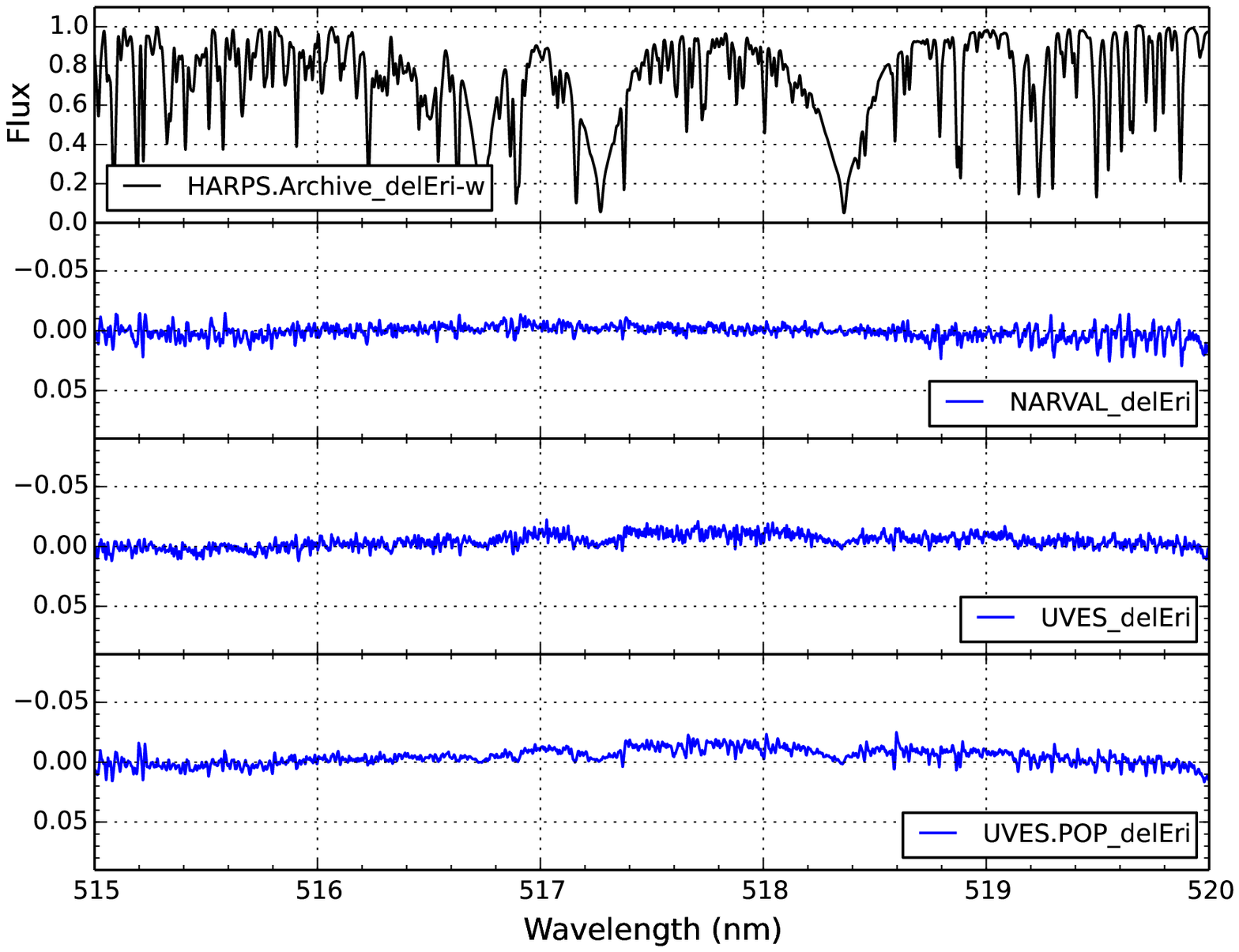}
        \includegraphics[width=9cm, scale=0.40,trim = 10mm 10mm 10mm 10mm, clip]{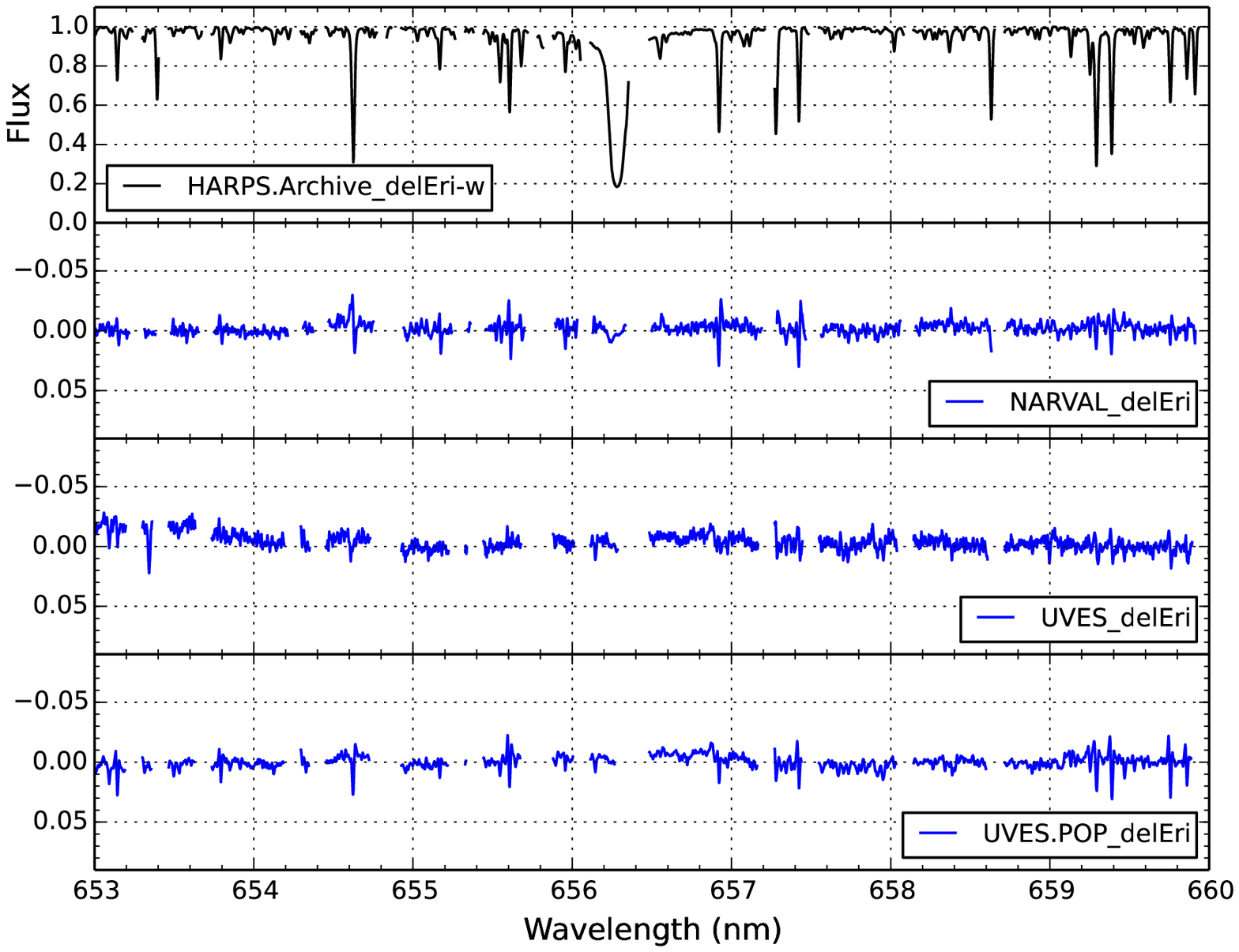}
        \includegraphics[width=9cm, scale=0.40,trim = 10mm 0mm 10mm 10mm, clip]{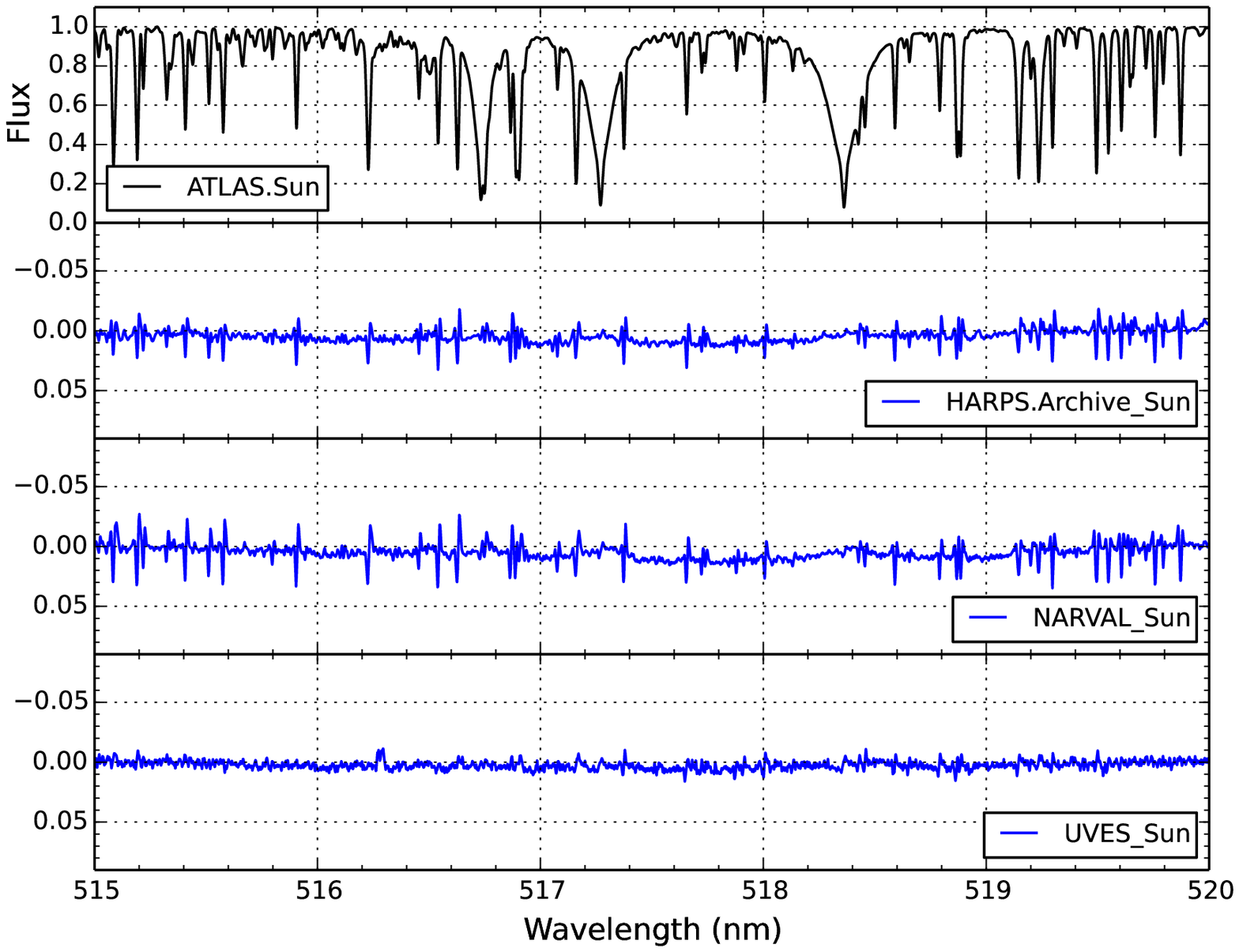}
        \includegraphics[width=9cm, scale=0.40,trim = 10mm 0mm 10mm 10mm, clip]{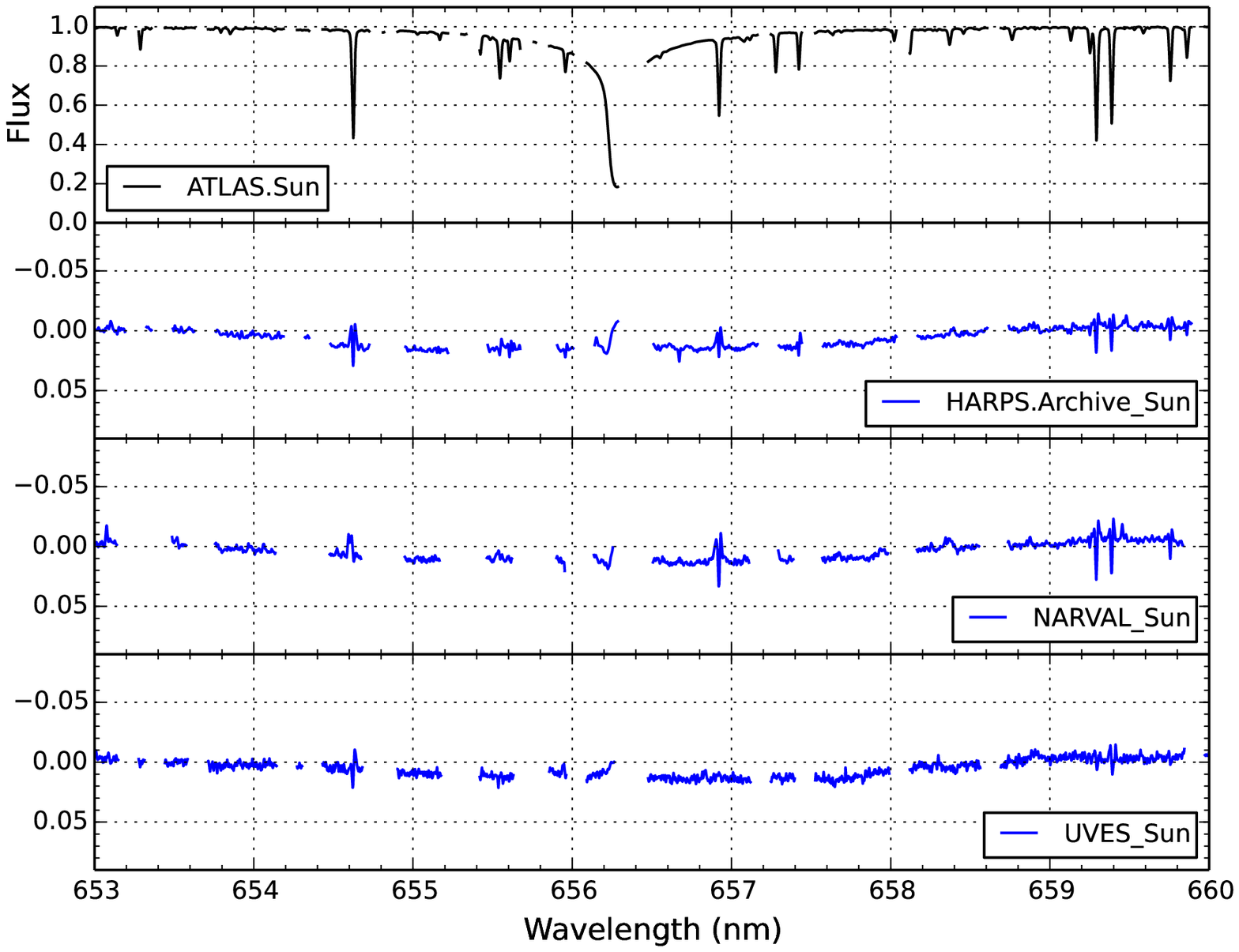}
        \par
    \end{centering}
    \caption{Normalization comparison for $\delta$ Eri (upper part) and the Sun (lower part). The spectra of reference are in black and the flux differences with the remaining observation are in blue. The regions compared correspond to the magnesium triplet (left) and H-$\alpha$ (right). Regions affected by tellurics were removed.}
    \label{fig:normalization_comparison}
\end{figure*}

No general trends as a function of wavelength or systematic effects were found even in difficult regions with strong lines such as H-$\alpha$ and the Mg triplet (see Fig.~\ref{fig:normalization_comparison}). The mean relative difference\footnote{The mean relative difference is defined by $\Delta\mathrm{flux} = \frac{\mathrm{flux}_{\mathrm{reference}}-\mathrm{flux}}{\mathrm{flux}_{\mathrm{reference}}}$} is 0.001$\pm$0.002 (0.1\%) and the average RMS is 0.009$\pm$0.004, which as expected depends on the type of the star (see Table ~\ref{tab:normalization}) since colder stars have more absorption lines and the continuum fitting process becomes more difficult.

\begin{table}
\begin{center}
\caption{Average RMS difference in flux for the normalized spectra of the same star but observed by different instruments.} \label{tab:normalization}
\tabcolsep=0.25cm
\begin{tabular}{c c c c c}
\hline
\textbf{Star} & \textbf{All} & \textbf{H-$\alpha$} & \textbf{H-$\beta$} & \textbf{MgTriplet} \\
\hline
$\alpha$ Cen A & 0.012 & 0.009 & 0.008 & 0.004 \\
$\alpha$ Cet & 0.021 & 0.015 & 0.026 & 0.018 \\
$\alpha$ Tau & 0.014 & 0.010 & 0.015 & 0.013 \\
$\beta$ Hyi & 0.009 & 0.009 & 0.008 & 0.003 \\
$\beta$ Vir & 0.007 & 0.007 & 0.006 & 0.005 \\
$\delta$ Eri & 0.008 & 0.006 & 0.010 & 0.007 \\
$\epsilon$ Eri & 0.015 & 0.015 & 0.008 & 0.007 \\
$\epsilon$ Vir & 0.006 & 0.005 & 0.006 & 0.006 \\
$\eta$ Boo & 0.005 & 0.005 & 0.005 & 0.004 \\
$\mu$ Ara & 0.012 & 0.014 & 0.011 & 0.008 \\
$\tau$ Cet & 0.007 & 0.008 & 0.007 & 0.005 \\
18 Sco & 0.007 & 0.009 & 0.007 & 0.007 \\
Arcturus & 0.010 & 0.010 & 0.012 & 0.015 \\
HD 107328 & 0.006 & 0.008 & 0.008 & 0.009 \\
HD 122563 & 0.008 & 0.007 & 0.007 & 0.004 \\
HD 140283 & 0.009 & 0.006 & 0.006 & 0.004 \\
HD 220009 & 0.006 & 0.005 & 0.007 & 0.007 \\
HD 22879 & 0.007 & 0.006 & 0.007 & 0.005 \\
HD 84937 & 0.007 & 0.005 & 0.009 & 0.005 \\
Procyon & 0.013 & 0.008 & 0.008 & 0.005 \\
Sun & 0.008 & 0.004 & 0.008 & 0.005 \\
\hline
Mean & 0.009 & 0.008 & 0.009 & 0.007 \\
StdDev & 0.004 & 0.003 & 0.004 & 0.004 \\
\hline
\end{tabular}
\tablefoot{Apart from the whole wavelength range (All), three individual regions are also reported: H-$\alpha$ (653 - 660 nm), H-$\beta$ (483 - 489 nm), and Mg triplet (515 - 520 nm).}
\end{center}
\end{table}

%\begin{figure*}
    %\begin{centering}
        %\includegraphics[width=9cm, scale=0.40,trim = 10mm 10mm 10mm 10mm, clip]{Images/{normalization_hbeta_HARPS.Archive_delEri-w}.png}
        %\includegraphics[width=9cm, scale=0.40,trim = 10mm 10mm 10mm 10mm, clip]{Images/{normalization_mgtriplet_HARPS.Archive_delEri-w}.png}
        %\includegraphics[width=9cm, scale=0.40,trim = 10mm 10mm 10mm 10mm, clip]{Images/{normalization_halpha_HARPS.Archive_delEri-w}.png}
        %\includegraphics[width=9cm, scale=0.40,trim = 10mm 10mm 10mm 10mm, clip]{Images/{normalization_whole_HARPS.Archive_delEri-w}.png}
        %\par
    %\end{centering}
    %\caption{Normalization comparison for $\delta$ Eri, which was observed by all the instruments. The regions compared correspond to: H-alpha (lower left), H-Beta (upper left), Magnesium triplet (upper right) and the whole spectra range (lower right).}
    %\label{fig:normalization_comparison_delEri}
%\end{figure*}

Additionally, we compared our normalized spectra with the atlas spectra (Sun and Arcturus), which were normalized by an independent external procedure. The mean relative difference is -0.001$\pm$0.002 (-0.1\%) for the Sun and -0.003\%$\pm$0.002\% (-0.3\%) for Arcturus. The average RMS is smaller than 0.02 (see Table ~\ref{tab:normalization_ATLAS}) and no systematic effects were found. The normalization in strong line regions is also consistent (see Fig.~\ref{fig:normalization_comparison}).

\begin{table}
\begin{center}
\caption{Mean RMS differences in flux for atlas spectra and our normalized spectra observed by different instruments.} \label{tab:normalization_ATLAS}
\tabcolsep=0.30cm
\begin{tabular}{c c c c c}
\hline
\textbf{Star} & \textbf{All} & \textbf{H-$\alpha$} & \textbf{H-$\beta$} & \textbf{MgTriplet} \\
\hline
Arcturus & 0.015 & 0.023 & 0.022 & 0.018 \\
Sun & 0.009 & 0.009 & 0.018 & 0.007 \\
\hline
\end{tabular}
\end{center}
\end{table}

%\begin{figure*}
    %\begin{centering}
        %\includegraphics[width=9cm, scale=0.40,trim = 10mm 10mm 10mm 10mm, clip]{Images/{normalization_hbeta_ATLAS.Sun}.png}
        %\includegraphics[width=9cm, scale=0.40,trim = 10mm 10mm 10mm 10mm, clip]{Images/{normalization_mgtriplet_ATLAS.Sun}.png}
        %\includegraphics[width=9cm, scale=0.40,trim = 10mm 10mm 10mm 10mm, clip]{Images/{normalization_halpha_ATLAS.Sun}.png}
        %\includegraphics[width=9cm, scale=0.40,trim = 10mm 10mm 10mm 10mm, clip]{Images/{normalization_whole_ATLAS.Sun}.png}
        %\par
    %\end{centering}
    %\caption{Normalization comparison for solar spectra. The regions compared are: H-alpha (lower left), H-Beta (upper left), Magnesium triplet (upper right) and the whole spectra range (lower right).}
    %\label{fig:normalization_comparison_ATLAS}
%\end{figure*}

\subsection{Resolution}
\label{sub:resolution}

In order to assess the quality of a spectrum, it is important to validate that the assumed initial resolving power is correct and that the convolution process works correctly.

We assumed that the final resulting library has a resolution of 70,000, which corresponds to a FWHM of $4.28$~km/s ($\mathrm{FWHM} = \frac{c}{\mathrm{R}}$). By using only the regions affected by the telluric lines, we cross-correlated the original and convolved normalized spectrum with the telluric mask and we obtained the FWHM in both cases ($FWHM_{1}$ and $FWHM_{2}$ respectively). Therefore, we can quantify how much the FWHM was changed as a result of the convolution process and validate that the difference is consistent with the expected initial resolution, that is,

\begin{equation}
    \label{eq:delta_fwhm}
    \Delta\mathrm{FWHM_{cross}} = (4.28 + (\mathrm{FWHM}_{1} - \mathrm{FWHM}_{2})) - (c / \mathrm{R}_{initial}),
\end{equation}

where $c$ is the speed of light and $R_{initial}$ the expected initial resolution (see Tables~\ref{tab:narval}, \ref{tab:harps}, \ref{tab:uves}, and \ref{tab:uves-pop}), should be close to zero.

The results\footnote{Atlas spectra were not considered since the telluric lines are not present in the original spectra.} presented in Table ~\ref{tab:resolution_validation} are in general agreement with the expected original resolving power. UVES spectra show the biggest dispersion in measured original resolving power, which is natural since not all the original spectra share the exact same resolving power (it can even vary between the lower and higher parts of the spectrum). 

\begin{table}
\begin{center}
\caption{Mean difference between expected and measured FWHM in km/s (see Eq.~\ref{eq:delta_fwhm}) and mean estimated original resolving power with their corresponding standard deviation.} \label{tab:resolution_validation}
\tabcolsep=0.55cm
\begin{tabular}{c c c}
\hline
\textbf{Instrument} & \textbf{$\Delta$FWHM$_{\mathrm{cross}}$ }  & \textbf{Estimated R$_{\mathrm{initial}}$} \\
 & $\mu\pm\sigma$ & $\mu\pm\sigma$ \\
\hline
HARPS & -0.07$\pm$0.21 & 118825$\pm$9655 \\
NARVAL & 0.10$\pm$0.13 & 79050$\pm$2999 \\
UVES & 0.19$\pm$0.14 & 83200$\pm$10397 \\
UVES.POP & -0.01$\pm$0.15 & 80932$\pm$3942 \\
\hline
\end{tabular}
\end{center}
\end{table}

We recall that NARVAL resolving power is not constant along the entire wavelength range. To show that this is not problematic, we measured the individual FWHM of a group of stellar lines (see Sect.~\ref{sub:EW}) for each spectra. For those stars with more than one spectrum, we calculated the relative difference in FWHM for the lines in common (see Table ~\ref{tab:fwhm_ew_comparison}):

\begin{equation}
    \Delta\mathrm{FWHM} = \frac{\mathrm{FWHM_{other}} - \mathrm{FWHM_{HARPS}}}{FWHM_{HARPS}} \mathrm{.}
\end{equation}

\begin{table}
\begin{center}
\caption{Difference in FWHM and equivalent widths between HARPS observations and the other instruments.} \label{tab:fwhm_ew_comparison}
\begin{tabular}{c c c c}
\hline
\textbf{Star} & \textbf{Instrument}  & \textbf{$\Delta$FWHM} & \textbf{$\Delta$EW} \\
 & & $\mu\pm\sigma$ & $\mu\pm\sigma$ \\
\hline

Sun & NARVAL & -0.08$\pm$0.05 & -0.00$\pm$0.03 \\
Sun & UVES & -0.05$\pm$0.04 & -0.02$\pm$0.06 \\
Arcturus & NARVAL & -0.06$\pm$0.03 & 0.02$\pm$0.02 \\
Arcturus & UVES & 0.00$\pm$0.03 & 0.03$\pm$0.03 \\
Arcturus & UVES.POP & -0.05$\pm$0.03 & 0.02$\pm$0.03 \\
18 Sco & NARVAL & -0.00$\pm$0.07 & 0.02$\pm$0.07 \\
$\delta$ Eri & NARVAL & -0.02$\pm$0.05 & -0.01$\pm$0.05 \\
$\delta$ Eri & UVES & 0.01$\pm$0.05 & 0.00$\pm$0.06 \\
$\delta$ Eri & UVES.POP & -0.02$\pm$0.05 & 0.00$\pm$0.05 \\
$\eta$ Boo & NARVAL & 0.00$\pm$0.04 & 0.00$\pm$0.05 \\
HD 220009 & NARVAL & -0.02$\pm$0.05 & -0.01$\pm$0.06 \\
Procyon & NARVAL & -0.01$\pm$0.05 & -0.00$\pm$0.06 \\
Procyon & UVES & 0.03$\pm$0.07 & 0.01$\pm$0.09 \\
Procyon & UVES.POP & 0.02$\pm$0.05 & 0.02$\pm$0.11 \\
\hline
Mean &  & -0.02$\pm$0.03 & 0.01$\pm$0.01 \\
\hline
\end{tabular}
\end{center}
\end{table}

\begin{figure*}
    \begin{centering}
        \includegraphics[width=9cm, scale=0.40,trim = 10mm 0mm 10mm 10mm, clip]{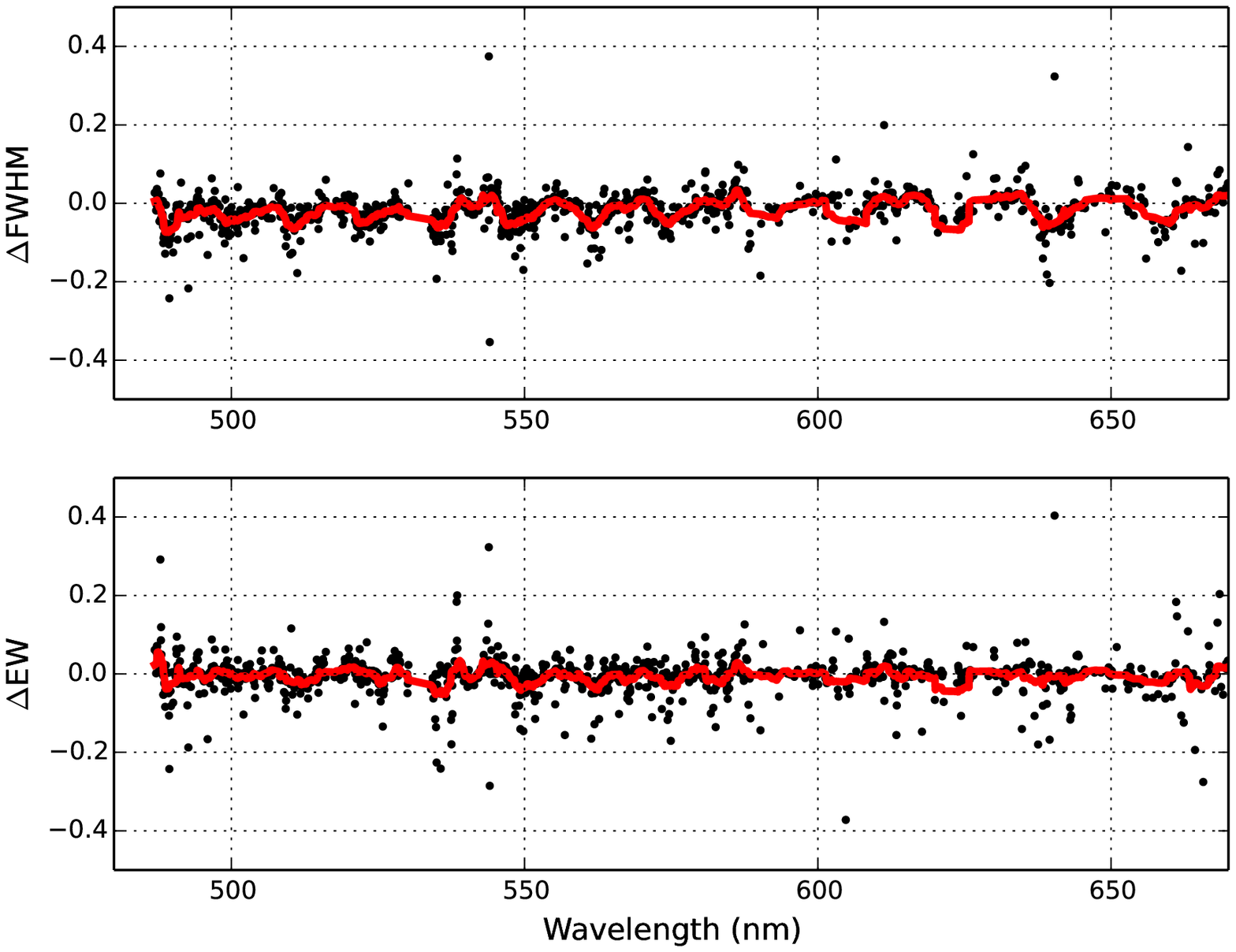}
       \includegraphics[width=9cm, scale=0.40,trim = 10mm 2mm 10mm 10mm, clip]{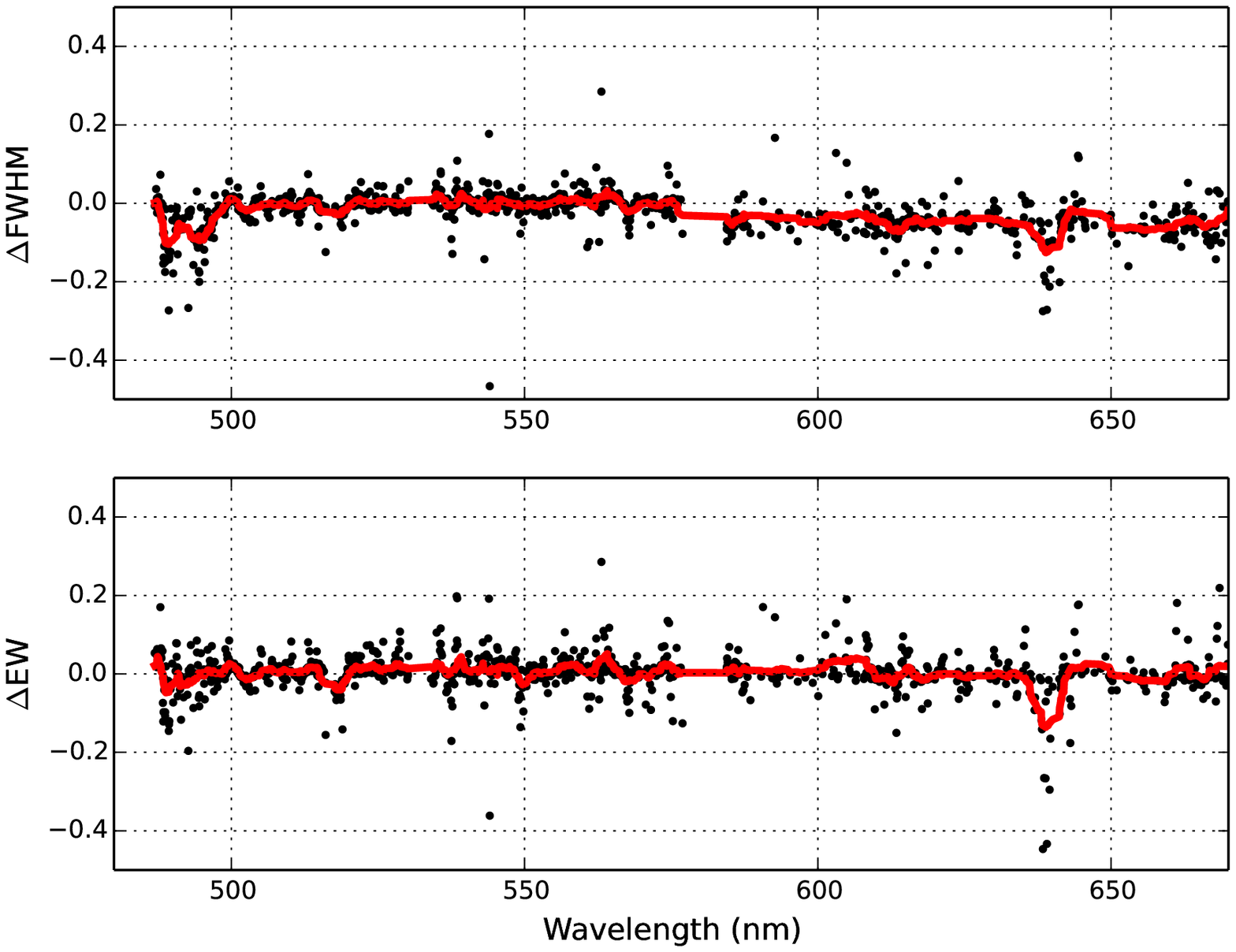}
        \par
    \end{centering}
    \caption{Relative difference in FWHM (upper part) and equivalent widths (lower part) for the $\delta$ Eri spectrum, with the moving average overplotted (red) for visual guidance. The numeric details of the HARPS against NARVAL (left plot) and HARPS against UVES-POP (right plot) comparisons can be found in Table ~\ref{tab:fwhm_ew_comparison}.}
    \label{fig:fwhm_ew_comparison}
\end{figure*}

The relative difference in FWHM for a spectrum observed by HARPS and NARVAL compared to the same spectrum observed by HARPS and UVES is not significant as we show in Fig.~\ref{fig:fwhm_ew_comparison}. The variable resolving power of NARVAL is small enough to be neglected and it does not have a relevant impact in spectroscopic analyses, such as the one performed in Paper~III.

\subsection{Radial velocity}
\label{sub:valid_rv}

The HARPS pipeline provides high precision radial velocity measurements for each observed spectrum. We used those measurements to test our radial velocity determination method and we obtained a zero mean difference with a standard deviation of $0.08$. The individual differences can be found in Table ~\ref{tab:harps}.

\subsection{Equivalent widths}
\label{sub:EW}

Different observations of the same Benchmark Star with the same resolution should have the same equivalent widths (EW). Thus, measuring and comparing EW provides us a different perspective to validate the normalization and convolution processes.

In that sense, we used the lines from \cite{2011ApJ...743..135R} for the Sun and Arcturus, which are good representatives for the dwarf and giant stellar types. The authors derived the EWs from high resolution spectra (R>300,000\footnote{\url{http://kurucz.harvard.edu/sun/fluxatlas/fluxatlastext.tab}} for the Sun, \cite{1984sfat.book.....K}, R$\sim$100,000 for Arcturus \citep{2005ASPC..336..321H}), therefore we used the original non-convolved observations of the Benchmarks Stars for the comparison.

The analysis was complemented with the inclusion of lines measured by \cite{2006AJ....131.3069L} and \cite{2007AJ....133.2464L} for 11 additional stars (seven dwarfs, four giants). In this case, the authors measured the EWs from spectra with a resolving power of 60,000. In consequence, we downgraded the resolution of the corresponding stars to match the same value and thereby equalize the conditions (i.e., same degree of blends).

Our EWs are determined by fitting a Gaussian profile in each absorption line and integrating its area. In the process, we discarded those lines that do not present a good fit (i.e., RMS > 0.05). The relative EW difference is calculated as:

\begin{equation} 
    \Delta\mathrm{EW} = \frac{\mathrm{EW}_{external} - \mathrm{EW}}{EW} \mathrm{.}
\end{equation}

\subsubsection{External consistency}

\begin{figure}
    \includegraphics[width=\linewidth, trim = 5mm 0mm 10mm 10mm]{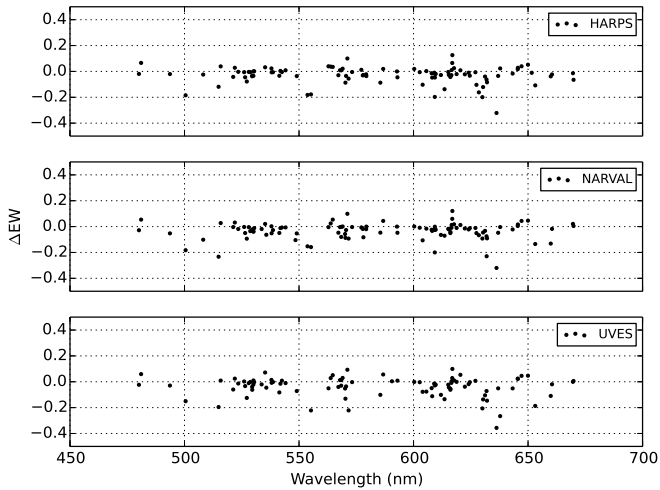}
    \caption{Relative differences in the equivalent width measured for the Sun by \cite{2011ApJ...743..135R}.}
    \label{fig:external_EW_comparison_sun}
\end{figure}

\begin{figure}
    \includegraphics[width=\linewidth, trim = 5mm 0mm 10mm 10mm]{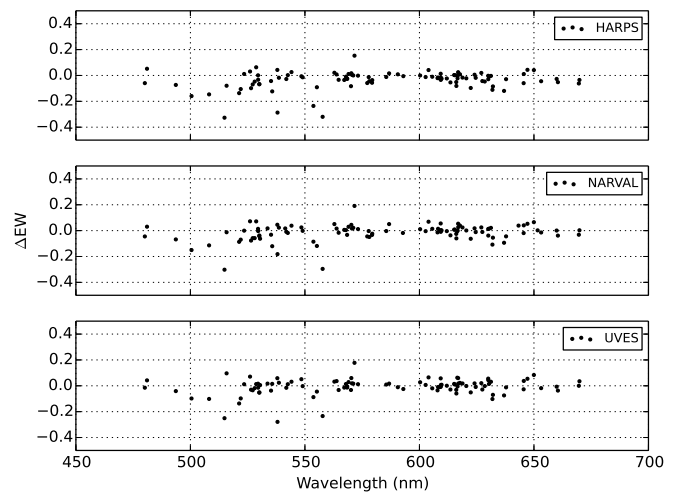}
    \caption{Relative differences in the equivalent width measured for Arcturus by \cite{2011ApJ...743..135R}.}
    \label{fig:external_EW_comparison_arcturus}
\end{figure}

We obtained a high level of consistency with the independently measured EWs (see Fig.~\ref{fig:external_EW_comparison_sun} and Fig.~\ref{fig:external_EW_comparison_arcturus}), the mean relative EW difference is around $-0.03\pm0.07$ for the Sun and $-0.03\pm0.12$ for Arcturus.  The dispersion is logically higher for the latter since the continuum in giants is less trivial to fit.

\begin{figure}
    \includegraphics[width=\linewidth, trim = 5mm 0mm 5mm 10mm]{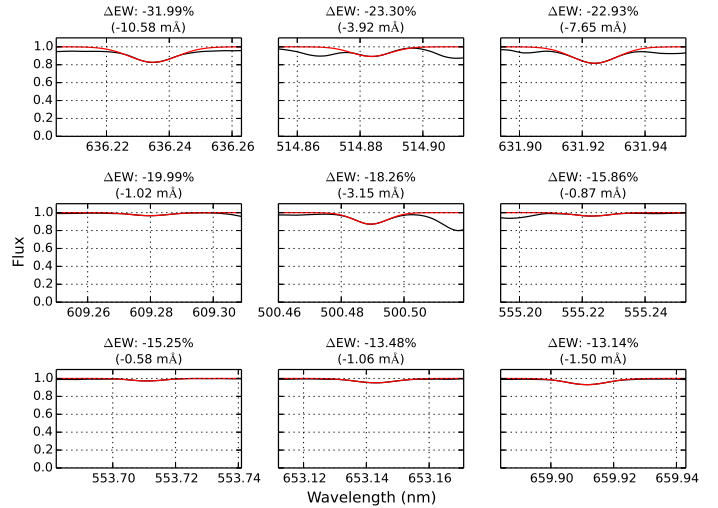}
    \caption{Normalized fluxes (black) and fitted Gaussians (red) for the lines with the highest relative difference in EW compared to \cite{2011ApJ...743..135R}. The fluxes correpond to the co-added spectra of the Sun observed by NARVAL.}
    \label{fig:worst_external_EW_sun}
\end{figure}

\begin{figure}
    \includegraphics[width=\linewidth, trim = 5mm 0mm 5mm 10mm]{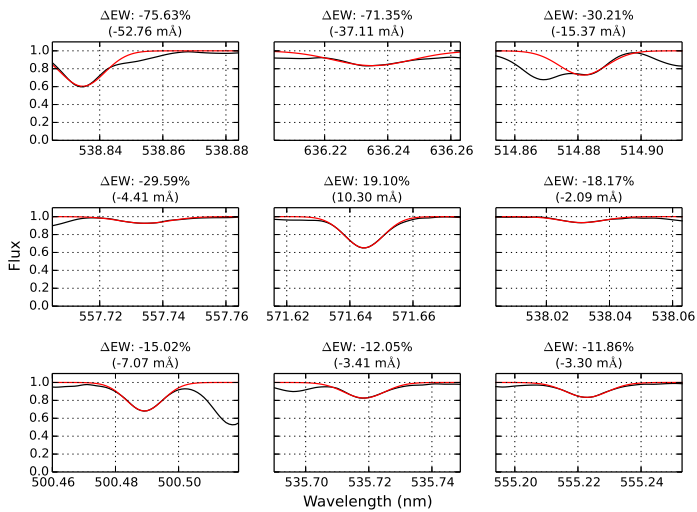}
    \caption{Normalized fluxes (black) and fitted Gaussians (red) for the lines with the highest relative difference in EW compared to \cite{2011ApJ...743..135R}. The fluxes correpond to the Arcturus spectrum observed by NARVAL.}
    \label{fig:worst_external_EW_arcturus}
\end{figure}

The visual inspection of lines with higher relative EW differences (see Fig.~\ref{fig:worst_external_EW_sun} and Fig.~\ref{fig:worst_external_EW_arcturus}) shows that our fit is consistent. The cause of the differences seems to be related to the continuum placement, which is specially amplified for weak absorption lines.

The additional stars present the same high level of agreement with a mean relative difference of $-0.01\pm0.04$ (1\%), all the detailed results can be found in Table ~\ref{tab:external_ew_comparison}.

\begin{table}
\begin{center}
\caption{Relative EW differences with external measurements \citep{2011ApJ...743..135R, 2006AJ....131.3069L, 2007AJ....133.2464L}.} \label{tab:external_ew_comparison}
\begin{tabular}{c c c c}
\hline
\textbf{Star} & \textbf{Instrument}  &  \textbf{$\Delta$EW} & \textbf{Num. of lines} \\
 & & $\mu\pm\sigma$ & \\
\hline
Sun & Atlas & 0.01$\pm$0.07 & 107 \\
Sun & HARPS & -0.03$\pm$0.07 & 101 \\
Sun & NARVAL & -0.04$\pm$0.06 & 106 \\
Sun & UVES & -0.04$\pm$0.08 & 94 \\
Arcturus & Atlas & -0.02$\pm$0.12 & 109 \\
Arcturus & HARPS & -0.05$\pm$0.12 & 104 \\
Arcturus & NARVAL & -0.03$\pm$0.12 & 105 \\
Arcturus & UVES & -0.02$\pm$0.12 & 101 \\
Arcturus & UVES.POP & -0.03$\pm$0.12 & 100 \\
18 Sco & HARPS & -0.02$\pm$0.15 & 866 \\
18 Sco & NARVAL & 0.00$\pm$0.15 & 885 \\
$\delta$ Eri & HARPS & 0.01$\pm$0.15 & 1071 \\
$\delta$ Eri & NARVAL & 0.01$\pm$0.15 & 1109 \\
$\delta$ Eri & UVES & 0.02$\pm$0.16 & 1055 \\
$\delta$ Eri & UVES.POP & 0.02$\pm$0.15 & 1072 \\
$\beta$ Gem & HARPS & 0.02$\pm$0.16 & 1176 \\
$\eta$ Boo & HARPS & -0.04$\pm$0.14 & 562 \\
$\eta$ Boo & NARVAL & -0.04$\pm$0.14 & 564 \\
HD 220009 & HARPS & 0.00$\pm$0.15 & 1048 \\
HD 220009 & NARVAL & -0.00$\pm$0.14 & 1075 \\
Procyon & HARPS & -0.07$\pm$0.12 & 530 \\
Procyon & NARVAL & -0.07$\pm$0.11 & 548 \\
Procyon & UVES & -0.06$\pm$0.13 & 524 \\
Procyon & UVES.POP & -0.05$\pm$0.12 & 526 \\
61 Cyg A & NARVAL & 0.04$\pm$0.23 & 717 \\
61 Cyg B & NARVAL & 0.12$\pm$0.31 & 764 \\
Gmb 1830 & NARVAL & 0.06$\pm$0.21 & 414 \\
$\mu$ Cas & NARVAL & -0.00$\pm$0.15 & 675 \\
$\mu$ Leo & NARVAL & 0.03$\pm$0.20 & 1041 \\
\hline
\multicolumn{2}{c}{Mean} & -0.01$\pm$0.04 \\
\hline
\end{tabular}
\end{center}
\end{table}

\subsubsection{Internal consistency}

We compared the measured EWs from those stars observed by different instruments. The level of internal consistency is very high (see Fig.~\ref{fig:fwhm_ew_comparison}) and the mean relative EW difference is $0.01\pm0.01$ (1\%) as shown in Table ~\ref{tab:fwhm_ew_comparison}. We estimated that abundance analysis based on EW methods show a very small variation of the order of $\pm$0.007 dex in metallicity when EWs are changed by 1\% (based on the analysis of a solar spectrum).

\section{Resulting library}
\label{sub:results}

The latest version of the library can be downloaded from \url{http://www.blancocuaresma.com/s/}. In this section we describe the contained data and their file formats.

\subsection{Spectra}                                                          

The library contains 78 spectra corresponding to the 34 benchmark stars. They cover the spectral range from 480 to 680~nm, homogeneously sampled with wavelength step of 0.001~nm (equivalent to 190,000 bins).

We provide four library variants: convolved/not convolved original non-normalized fluxes and convolved/not convolved normalized spectra.

The spectra are saved in two different formats:

\begin{enumerate}
    \item FITS format, following the standards of the IAU\footnote{International Astronomical Union} \citep{2002A&A...395.1061G, 2006A&A...446..747G}, where the spectral coordinates (wavelengths) are specified in the header via CRVAL1 and CDELT1 keywords. The FITS headers also contain metadata for each spectrum, such as their observation date, instrument, celestial coordinates, and history log. The fluxes and errors are stored, respectively, in the primary data unit and in an image extension \citep{1988A&AS...73..359G} as 1D arrays.
    \item Compressed plain text files with three columns delimited by tabulations: wavelength (nm), flux, and error.
\end{enumerate}

\subsection{Telluric lines}                                    

As described in Sect.~\ref{sub:tellurics}, regions potentially affected by telluric lines were identified, thus the user can discard them easily. For each spectrum, we provide a plain text file with three columns (delimited by tabulations) which correspond to the telluric line peak, beginning, and end of the affected region (in nanometers).

\subsubsection{Gap regions}                                                         
                                                                                 
As described in Sect.~\ref{sub:gaps}, some spectra contain gaps (regions without valid fluxes). We provide those regions in individual plain text files (one per spectrum) with two columns (delimited by tabulations), which correspond to the beginning and end of the gap (in nanometers).

\section{Conclusions}
\label{sub:conclusions}

We created a homogeneous library of high resolution and high S/N spectra corresponding to 34 benchmark stars with four different variants (convolved/not convolved original non-normalized fluxes, convolved/not convolved normalized spectra). The library provides a powerful tool to assess spectral analysis methods and calibrate spectroscopic surveys.

We validated the consistency of the library by carefully checking the normalization and convolution treatments. The radial velocity corrections was certified by comparing the results with the high precision measurements of HARPS pipeline. We verified the coherence of the treated spectra by comparing them with EW measurements completely independent from our process. These strict tests proved the high quality level of the spectral library. 

The whole creation and verification process was automatized, minimizing human subjectivity and ensuring reproducibility. It also allows us to create new versions of the library adapted to particular needs (i.e., different resolutions and spectral ranges) of specific spectroscopic surveys or spectral analyses.

The Gaia FGK benchmark stars library provides an opportunity to homogenize spectroscopic results (from single observations to massive surveys), reducing their dispersion and making them more comparable. This higher level of homogeneity can lead to a better and more robust understanding of the Galaxy such as its formation, evolution, and current structure.

\begin{acknowledgements}
    We thank N. Brouillet and T. Jacq for the help in the library construction. We also thank G. Sacco for the ingestion of our library in the second data release of the Gaia-ESO Survey.
    This work was partially supported by the Gaia Research for European Astronomy Training (GREAT-ITN) Marie Curie network, funded through the European Union Seventh Framework Programme [FP7/2007-2013] under grant agreement n. 264895.
    UH acknowledges support from the Swedish National Space Board (Rymdstyrelsen).
    The authors acknowledge the role of the SAM collaboration (\url{http://www.astro.uu.se/~ulrike/GaiaSAM}) in stimulating this research through regular workshops.
    Computer time for this study was provided by the computing facilities MCIA (M\'esocentre de Calcul Intensif Aquitain) of the Universit\'e de Bordeaux and of the Universit\'e de Pau et des Pays de l'Adour.
    All the software used in the data analysis were provided by the Open Source community.
\end{acknowledgements}

%-------------------------------------------------------------------

\bibliographystyle{bibtex/aa} % style aa.bst
\bibliography{References} % your references Yourfile.bib

\begin{thebibliography}{53}
\expandafter\ifx\csname natexlab\endcsname\relax\def\natexlab#1{#1}\fi

\bibitem[{{Allende Prieto}(2007)}]{2007AJ....134.1843A}
{Allende Prieto}, C. 2007, \aj, 134, 1843

\bibitem[{{Allende Prieto} {et~al.}(2008{\natexlab{a}}){Allende Prieto},
  {Majewski}, {Schiavon}, {Cunha}, {Frinchaboy}, {Holtzman}, {Johnston},
  {Shetrone}, {Skrutskie}, {Smith}, \& {Wilson}}]{2008AN....329.1018A}
{Allende Prieto}, C., {Majewski}, S.~R., {Schiavon}, R., {et~al.}
  2008{\natexlab{a}}, Astronomische Nachrichten, 329, 1018

\bibitem[{{Allende Prieto} {et~al.}(2008{\natexlab{b}}){Allende Prieto},
  {Sivarani}, {Beers}, {Lee}, {Koesterke}, {Shetrone}, {Sneden}, {Lambert},
  {Wilhelm}, {Rockosi}, {Lai}, {Yanny}, {Ivans}, {Johnson}, {Aoki},
  {Bailer-Jones}, \& {Re Fiorentin}}]{2008AJ....136.2070A}
{Allende Prieto}, C., {Sivarani}, T., {Beers}, T.~C., {et~al.}
  2008{\natexlab{b}}, \aj, 136, 2070

\bibitem[{{Auri{\`e}re}(2003)}]{2003EAS.....9..105A}
{Auri{\`e}re}, M. 2003, in EAS Publications Series, Vol.~9, EAS Publications
  Series, ed. J.~{Arnaud} \& N.~{Meunier}, 105

\bibitem[{{Ayres}(2010)}]{2010ApJS..187..149A}
{Ayres}, T.~R. 2010, \apjs, 187, 149

\bibitem[{{Bagnulo} {et~al.}(2003){Bagnulo}, {Jehin}, {Ledoux}, {Cabanac},
  {Melo}, {Gilmozzi}, \& {ESO Paranal Science Operations
  Team}}]{2003Msngr.114...10B}
{Bagnulo}, S., {Jehin}, E., {Ledoux}, C., {et~al.} 2003, The Messenger, 114, 10

\bibitem[{{Ballester} {et~al.}(2000){Ballester}, {Modigliani}, {Boitquin},
  {Cristiani}, {Hanuschik}, {Kaufer}, \& {Wolf}}]{2000Msngr.101...31B}
{Ballester}, P., {Modigliani}, A., {Boitquin}, O., {et~al.} 2000, The
  Messenger, 101, 31

\bibitem[{{Baranne} {et~al.}(1996){Baranne}, {Queloz}, {Mayor}, {Adrianzyk},
  {Knispel}, {Kohler}, {Lacroix}, {Meunier}, {Rimbaud}, \&
  {Vin}}]{1996A&AS..119..373B}
{Baranne}, A., {Queloz}, D., {Mayor}, M., {et~al.} 1996, \aaps, 119, 373

\bibitem[{{Bertaux} {et~al.}(2013){Bertaux}, {Lallement}, {Ferron}, {Boone}, \&
  {Bodichon}}]{2013arXiv1311.4169B}
{Bertaux}, J.~L., {Lallement}, R., {Ferron}, S., {Boone}, C., \& {Bodichon}, R.
  2013, ArXiv e-prints

\bibitem[{{Blanco-Cuaresma} {et~al.}(2013){Blanco-Cuaresma}, {Soubiran},
  {Jofr{\'e}}, \& {Heiter}}]{2013arXiv1312.4545B}
{Blanco-Cuaresma}, S., {Soubiran}, C., {Jofr{\'e}}, P., \& {Heiter}, U. 2013,
  ArXiv e-prints

\bibitem[{{Dekker} {et~al.}(2000){Dekker}, {D'Odorico}, {Kaufer}, {Delabre}, \&
  {Kotzlowski}}]{2000SPIE.4008..534D}
{Dekker}, H., {D'Odorico}, S., {Kaufer}, A., {Delabre}, B., \& {Kotzlowski}, H.
  2000, in Society of Photo-Optical Instrumentation Engineers (SPIE) Conference
  Series, Vol. 4008, Society of Photo-Optical Instrumentation Engineers (SPIE)
  Conference Series, ed. M.~{Iye} \& A.~F. {Moorwood}, 534--545

\bibitem[{{Donati} {et~al.}(1997){Donati}, {Semel}, {Carter}, {Rees}, \&
  {Collier Cameron}}]{donati97}
{Donati}, J., {Semel}, M., {Carter}, B.~D., {Rees}, D.~E., \& {Collier
  Cameron}, A. 1997, \mnras, 291, 658

\bibitem[{{Freeman}(2010)}]{2010gama.conf..319F}
{Freeman}, K.~C. 2010, in Galaxies and their Masks, ed. D.~L. {Block}, K.~C.
  {Freeman}, \& I.~{Puerari}, 319

\bibitem[{{Gilmore} {et~al.}(2012){Gilmore}, {Randich}, {Asplund}, {Binney},
  {Bonifacio}, {Drew}, {Feltzing}, {Ferguson}, {Jeffries}, {Micela},
  {Negueruela}, {Prusti}, {Rix}, {Vallenari}, {Alfaro}, {Allende-Prieto},
  {Babusiaux}, {Bensby}, {Blomme}, {Bragaglia}, {Flaccomio}, {Fran{\c c}ois},
  {Irwin}, {Koposov}, {Korn}, {Lanzafame}, {Pancino}, {Paunzen},
  {Recio-Blanco}, {Sacco}, {Smiljanic}, {Van Eck}, \&
  {Walton}}]{2012Msngr.147...25G}
{Gilmore}, G., {Randich}, S., {Asplund}, M., {et~al.} 2012, The Messenger, 147,
  25

\bibitem[{{Gray} \& {Corbally}(1994)}]{1994AJ....107..742G}
{Gray}, R.~O. \& {Corbally}, C.~J. 1994, \aj, 107, 742

\bibitem[{{Greisen} \& {Calabretta}(2002)}]{2002A&A...395.1061G}
{Greisen}, E.~W. \& {Calabretta}, M.~R. 2002, \aap, 395, 1061

\bibitem[{{Greisen} {et~al.}(2006){Greisen}, {Calabretta}, {Valdes}, \&
  {Allen}}]{2006A&A...446..747G}
{Greisen}, E.~W., {Calabretta}, M.~R., {Valdes}, F.~G., \& {Allen}, S.~L. 2006,
  \aap, 446, 747

\bibitem[{{Grosbol} {et~al.}(1988){Grosbol}, {Harten}, {Greisen}, \&
  {Wells}}]{1988A&AS...73..359G}
{Grosbol}, P., {Harten}, R.~H., {Greisen}, E.~W., \& {Wells}, D.~C. 1988,
  \aaps, 73, 359

\bibitem[{{Gustafsson} {et~al.}(2008){Gustafsson}, {Edvardsson}, {Eriksson},
  {J{\o}rgensen}, {Nordlund}, \& {Plez}}]{2008A&A...486..951G}
{Gustafsson}, B., {Edvardsson}, B., {Eriksson}, K., {et~al.} 2008, \aap, 486,
  951

\bibitem[{{Hinkle} \& {Wallace}(2005)}]{2005ASPC..336..321H}
{Hinkle}, K. \& {Wallace}, L. 2005, in Astronomical Society of the Pacific
  Conference Series, Vol. 336, Cosmic Abundances as Records of Stellar
  Evolution and Nucleosynthesis, ed. T.~G. {Barnes}, III \& F.~N. {Bash}, 321

\bibitem[{{Hinkle} {et~al.}(2000){Hinkle}, {Wallace}, {Valenti}, \&
  {Harmer}}]{2000vnia.book.....H}
{Hinkle}, K., {Wallace}, L., {Valenti}, J., \& {Harmer}, D. 2000, {Visible and
  Near Infrared Atlas of the Arcturus Spectrum 3727-9300 A}

\bibitem[{{Jofr\'e} {et~al.}(2013){Jofr\'e}, {Heiter}, {Soubiran},
  {Blanco-Cuaresma}, {Pancino}, {Bergemann}, {Cantat-Gaudin}, {Gonzalez
  Hernandez}, {Hill}, {Lardo}, {de Laverny}, {Lind}, {Magrini}, {Masseron},
  {Montes}, {Mucciarelli}, {Nordlander}, {Recio-Blanco}, {Sobeck}, {Sordo},
  {Sousa}, {Tabernero}, {Vallenari}, {Van Eck}, \&
  {Worley}}]{2013arXiv1309.1099J}
{Jofr\'e}, P., {Heiter}, U., {Soubiran}, C., {et~al.} 2013, ArXiv e-prints

\bibitem[{{Jofr{\'e}} {et~al.}(2010){Jofr{\'e}}, {Panter}, {Hansen}, \&
  {Weiss}}]{2010A&A...517A..57J}
{Jofr{\'e}}, P., {Panter}, B., {Hansen}, C.~J., \& {Weiss}, A. 2010, \aap, 517,
  A57

\bibitem[{{Katz} {et~al.}(1998){Katz}, {Soubiran}, {Cayrel}, {Adda}, \&
  {Cautain}}]{1998A&A...338..151K}
{Katz}, D., {Soubiran}, C., {Cayrel}, R., {Adda}, M., \& {Cautain}, R. 1998,
  \aap, 338, 151

\bibitem[{{Koleva} {et~al.}(2009){Koleva}, {Prugniel}, {Bouchard}, \&
  {Wu}}]{2009A&A...501.1269K}
{Koleva}, M., {Prugniel}, P., {Bouchard}, A., \& {Wu}, Y. 2009, \aap, 501, 1269

\bibitem[{{Kupka} {et~al.}(2011){Kupka}, {Dubernet}, \& {VAMDC
  Collaboration}}]{2011BaltA..20..503K}
{Kupka}, F., {Dubernet}, M.-L., \& {VAMDC Collaboration}. 2011, Baltic
  Astronomy, 20, 503

\bibitem[{{Kurucz} {et~al.}(1984){Kurucz}, {Furenlid}, {Brault}, \&
  {Testerman}}]{1984sfat.book.....K}
{Kurucz}, R.~L., {Furenlid}, I., {Brault}, J., \& {Testerman}, L. 1984, {Solar
  flux atlas from 296 to 1300 nm}

\bibitem[{{Lee} {et~al.}(2008){Lee}, {Beers}, {Sivarani}, {Allende Prieto},
  {Koesterke}, {Wilhelm}, {Re Fiorentin}, {Bailer-Jones}, {Norris}, {Rockosi},
  {Yanny}, {Newberg}, {Covey}, {Zhang}, \& {Luo}}]{2008AJ....136.2022L}
{Lee}, Y.~S., {Beers}, T.~C., {Sivarani}, T., {et~al.} 2008, \aj, 136, 2022

\bibitem[{{Luck} \& {Heiter}(2006)}]{2006AJ....131.3069L}
{Luck}, R.~E. \& {Heiter}, U. 2006, \aj, 131, 3069

\bibitem[{{Luck} \& {Heiter}(2007)}]{2007AJ....133.2464L}
{Luck}, R.~E. \& {Heiter}, U. 2007, \aj, 133, 2464

\bibitem[{{Magrini} {et~al.}(2013){Magrini}, {Randich}, {Friel}, {Spina},
  {Jacobson}, {Cantat-Gaudin}, {Donati}, {Baglioni}, {Maiorca}, {Bragaglia},
  {Sordo}, \& {Vallenari}}]{2013A&A...558A..38M}
{Magrini}, L., {Randich}, S., {Friel}, E., {et~al.} 2013, \aap, 558, A38

\bibitem[{{Mayor} {et~al.}(2003){Mayor}, {Pepe}, {Queloz}, {Bouchy},
  {Rupprecht}, {Lo Curto}, {Avila}, {Benz}, {Bertaux}, {Bonfils}, {Dall},
  {Dekker}, {Delabre}, {Eckert}, {Fleury}, {Gilliotte}, {Gojak}, {Guzman},
  {Kohler}, {Lizon}, {Longinotti}, {Lovis}, {Megevand}, {Pasquini}, {Reyes},
  {Sivan}, {Sosnowska}, {Soto}, {Udry}, {van Kesteren}, {Weber}, \&
  {Weilenmann}}]{2003Msngr.114...20M}
{Mayor}, M., {Pepe}, F., {Queloz}, D., {et~al.} 2003, The Messenger, 114, 20

\bibitem[{{Molaro} \& {Monai}(2012)}]{2012A&A...544A.125M}
{Molaro}, P. \& {Monai}, S. 2012, \aap, 544, A125

\bibitem[{{Mucciarelli} {et~al.}(2013){Mucciarelli}, {Pancino}, {Lovisi},
  {Ferraro}, \& {Lapenna}}]{2013ApJ...766...78M}
{Mucciarelli}, A., {Pancino}, E., {Lovisi}, L., {Ferraro}, F.~R., \& {Lapenna},
  E. 2013, \apj, 766, 78

\bibitem[{{Munari} \& {Sordo}(2005)}]{2005MSAIS...8..170M}
{Munari}, U. \& {Sordo}, R. 2005, Memorie della Societa Astronomica Italiana
  Supplementi, 8, 170

\bibitem[{{Pepe} {et~al.}(2002){Pepe}, {Mayor}, {Galland}, {Naef}, {Queloz},
  {Santos}, {Udry}, \& {Burnet}}]{2002A&A...388..632P}
{Pepe}, F., {Mayor}, M., {Galland}, F., {et~al.} 2002, \aap, 388, 632

\bibitem[{{Percival} {et~al.}(2009){Percival}, {Salaris}, {Cassisi}, \&
  {Pietrinferni}}]{2009ApJ...690..427P}
{Percival}, S.~M., {Salaris}, M., {Cassisi}, S., \& {Pietrinferni}, A. 2009,
  \apj, 690, 427

\bibitem[{{Perryman} {et~al.}(2001){Perryman}, {de Boer}, {Gilmore}, {H{\o}g},
  {Lattanzi}, {Lindegren}, {Luri}, {Mignard}, {Pace}, \& {de
  Zeeuw}}]{2001A&A...369..339P}
{Perryman}, M.~A.~C., {de Boer}, K.~S., {Gilmore}, G., {et~al.} 2001, A\&A,
  369, 339

\bibitem[{{Posbic} {et~al.}(2012){Posbic}, {Katz}, {Caffau}, {Bonifacio},
  {G{\'o}mez}, {Sbordone}, \& {Arenou}}]{2012A&A...544A.154P}
{Posbic}, H., {Katz}, D., {Caffau}, E., {et~al.} 2012, \aap, 544, A154

\bibitem[{{Prugniel} \& {Soubiran}(2001)}]{2001A&A...369.1048P}
{Prugniel}, P. \& {Soubiran}, C. 2001, \aap, 369, 1048

\bibitem[{{Ram{\'{\i}}rez} \& {Allende Prieto}(2011)}]{2011ApJ...743..135R}
{Ram{\'{\i}}rez}, I. \& {Allende Prieto}, C. 2011, \apj, 743, 135

\bibitem[{{Recio-Blanco} {et~al.}(2006){Recio-Blanco}, {Bijaoui}, \& {de
  Laverny}}]{2006MNRAS.370..141R}
{Recio-Blanco}, A., {Bijaoui}, A., \& {de Laverny}, P. 2006, \mnras, 370, 141

\bibitem[{{S{\'a}nchez-Bl{\'a}zquez} {et~al.}(2006){S{\'a}nchez-Bl{\'a}zquez},
  {Peletier}, {Jim{\'e}nez-Vicente}, {Cardiel}, {Cenarro},
  {Falc{\'o}n-Barroso}, {Gorgas}, {Selam}, \& {Vazdekis}}]{2006MNRAS.371..703S}
{S{\'a}nchez-Bl{\'a}zquez}, P., {Peletier}, R.~F., {Jim{\'e}nez-Vicente}, J.,
  {et~al.} 2006, \mnras, 371, 703

\bibitem[{{Steinmetz} {et~al.}(2006){Steinmetz}, {Zwitter}, {Siebert},
  {Watson}, {Freeman}, {Munari}, {Campbell}, {Williams}, {Seabroke}, {Wyse},
  {Parker}, {Bienaym{\'e}}, {Roeser}, {Gibson}, {Gilmore}, {Grebel}, {Helmi},
  {Navarro}, {Burton}, {Cass}, {Dawe}, {Fiegert}, {Hartley}, {Russell},
  {Saunders}, {Enke}, {Bailin}, {Binney}, {Bland-Hawthorn}, {Boeche}, {Dehnen},
  {Eisenstein}, {Evans}, {Fiorucci}, {Fulbright}, {Gerhard}, {Jauregi}, {Kelz},
  {Mijovi{\'c}}, {Minchev}, {Parmentier}, {Pe{\~n}arrubia}, {Quillen}, {Read},
  {Ruchti}, {Scholz}, {Siviero}, {Smith}, {Sordo}, {Veltz}, {Vidrih}, {von
  Berlepsch}, {Boyle}, \& {Schilbach}}]{2006AJ....132.1645S}
{Steinmetz}, M., {Zwitter}, T., {Siebert}, A., {et~al.} 2006, AJ, 132, 1645

\bibitem[{{Th{\'e}venin} {et~al.}(2005){Th{\'e}venin}, {Kervella}, {Pichon},
  {Morel}, {di Folco}, \& {Lebreton}}]{2005A&A...436..253T}
{Th{\'e}venin}, F., {Kervella}, P., {Pichon}, B., {et~al.} 2005, \aap, 436, 253

\bibitem[{{Valdes} {et~al.}(2004){Valdes}, {Gupta}, {Rose}, {Singh}, \&
  {Bell}}]{2004ApJS..152..251V}
{Valdes}, F., {Gupta}, R., {Rose}, J.~A., {Singh}, H.~P., \& {Bell}, D.~J.
  2004, \apjs, 152, 251

\bibitem[{{Valenti} \& {Piskunov}(1996)}]{1996A&AS..118..595V}
{Valenti}, J.~A. \& {Piskunov}, N. 1996, \aaps, 118, 595

\bibitem[{{Vazdekis} {et~al.}(2012){Vazdekis}, {Ricciardelli}, {Cenarro},
  {Rivero-Gonz{\'a}lez}, {D{\'{\i}}az-Garc{\'{\i}}a}, \&
  {Falc{\'o}n-Barroso}}]{2012MNRAS.424..157V}
{Vazdekis}, A., {Ricciardelli}, E., {Cenarro}, A.~J., {et~al.} 2012, \mnras,
  424, 157

\bibitem[{{Wu} {et~al.}(2011){Wu}, {Luo}, {Li}, {Shi}, {Prugniel}, {Liang},
  {Zhao}, {Zhang}, {Bai}, {Wei}, {Dong}, {Zhang}, \&
  {Chen}}]{2011RAA....11..924W}
{Wu}, Y., {Luo}, A.-L., {Li}, H.-N., {et~al.} 2011, Research in Astronomy and
  Astrophysics, 11, 924

\bibitem[{{York} {et~al.}(2000){York}, {Adelman}, {Anderson}, {Anderson},
  {Annis}, {Bahcall}, {Bakken}, {Barkhouser}, {Bastian}, {Berman}, {Boroski},
  {Bracker}, {Briegel}, {Briggs}, {Brinkmann}, {Brunner}, {Burles}, {Carey},
  {Carr}, {Castander}, {Chen}, {Colestock}, {Connolly}, {Crocker}, {Csabai},
  {Czarapata}, {Davis}, {Doi}, {Dombeck}, {Eisenstein}, {Ellman}, {Elms},
  {Evans}, {Fan}, {Federwitz}, {Fiscelli}, {Friedman}, {Frieman}, {Fukugita},
  {Gillespie}, {Gunn}, {Gurbani}, {de Haas}, {Haldeman}, {Harris}, {Hayes},
  {Heckman}, {Hennessy}, {Hindsley}, {Holm}, {Holmgren}, {Huang}, {Hull},
  {Husby}, {Ichikawa}, {Ichikawa}, {Ivezi{\'c}}, {Kent}, {Kim}, {Kinney},
  {Klaene}, {Kleinman}, {Kleinman}, {Knapp}, {Korienek}, {Kron}, {Kunszt},
  {Lamb}, {Lee}, {Leger}, {Limmongkol}, {Lindenmeyer}, {Long}, {Loomis},
  {Loveday}, {Lucinio}, {Lupton}, {MacKinnon}, {Mannery}, {Mantsch}, {Margon},
  {McGehee}, {McKay}, {Meiksin}, {Merelli}, {Monet}, {Munn}, {Narayanan},
  {Nash}, {Neilsen}, {Neswold}, {Newberg}, {Nichol}, {Nicinski}, {Nonino},
  {Okada}, {Okamura}, {Ostriker}, {Owen}, {Pauls}, {Peoples}, {Peterson},
  {Petravick}, {Pier}, {Pope}, {Pordes}, {Prosapio}, {Rechenmacher}, {Quinn},
  {Richards}, {Richmond}, {Rivetta}, {Rockosi}, {Ruthmansdorfer}, {Sandford},
  {Schlegel}, {Schneider}, {Sekiguchi}, {Sergey}, {Shimasaku}, {Siegmund},
  {Smee}, {Smith}, {Snedden}, {Stone}, {Stoughton}, {Strauss}, {Stubbs},
  {SubbaRao}, {Szalay}, {Szapudi}, {Szokoly}, {Thakar}, {Tremonti}, {Tucker},
  {Uomoto}, {Vanden Berk}, {Vogeley}, {Waddell}, {Wang}, {Watanabe},
  {Weinberg}, {Yanny}, {Yasuda}, \& {SDSS Collaboration}}]{2000AJ....120.1579Y}
{York}, D.~G., {Adelman}, J., {Anderson}, Jr., J.~E., {et~al.} 2000, \aj, 120,
  1579

\bibitem[{{Zhang} {et~al.}(2005){Zhang}, {Li}, \& {Han}}]{2005MNRAS.364..503Z}
{Zhang}, F., {Li}, L., \& {Han}, Z. 2005, \mnras, 364, 503

\bibitem[{{Zhao} {et~al.}(2006){Zhao}, {Chen}, {Shi}, {Liang}, {Hou}, {Chen},
  {Zhang}, \& {Li}}]{2006ChJAA...6..265Z}
{Zhao}, G., {Chen}, Y.-Q., {Shi}, J.-R., {et~al.} 2006, CJAA, 6, 265

\bibitem[{{Zucker}(2003)}]{2003MNRAS.342.1291Z}
{Zucker}, S. 2003, \mnras, 342, 1291

\end{thebibliography}

\end{document}